\pdfoutput=1
\documentclass[twocolumn]{aastex631}

\newcommand\lya{Ly$\alpha$}

\newcommand\mgii{Mg\,{\sc ii}}
\newcommand\civ{C\,{\sc iv}}
\newcommand\oii{O\,{\sc ii} 1305 \AA}
\newcommand\oiii{[O\,{\sc iii]} 4959,5007 \AA}
\newcommand\cii{C\,{\sc ii} 1335 \AA}
\newcommand\ip{$i_{775}$}
\newcommand\zp{$z_{850}$}
\newcommand\ciialma{[C\,{\sc ii}]\,158$\mu$m}

\shorttitle{Galaxies near a radio-loud quasar at $z=5.8$}
\shortauthors{Overzier}

\graphicspath{{./}{figures/}}
\begin{document}

\title{Conditions for direct black hole seed collapse near a radio-loud quasar 1 Gyr after the Big Bang}

\author[0000-0002-8214-7617]{Roderik A. Overzier}
\affiliation{Observat\'{o}rio Nacional/MCTIC, Rua General Jos\'{e} Cristino, 77, S\~{a}o Crist\'{o}v\~{a}o, Rio de Janeiro, RJ 20921-400, Brazil}
\affiliation{Institute of Astronomy, Geophysics and Atmospheric Sciences, University of S\~{a}o Paulo, Rua do Mat\~{a}o, 1226, S\~{a}o Paulo, SP 05508-090, Brazil}
\email{roderikoverzier@gmail.com}



\begin{abstract}
Observations of luminous quasars and their supermassive black holes at $z\gtrsim6$ suggest that they formed at dense matter peaks in the early universe. However, few studies have found definitive evidence that the quasars lie at cosmic density peaks, in clear contrast with theory predictions. Here we present new evidence that the radio-loud quasar SDSS J0836+0054 at $z=5.8$ could be part of a surprisingly rich structure of galaxies. This conclusion is reached by combining a number of findings previously reported in the literature. \citet{bosman20} obtained the redshifts of three companion galaxies, confirming an overdensity of \ip-dropouts found by \citet{zheng06}. By comparing this structure with those found near other quasars and large overdense regions in the field at $z\sim6-7$, we show that the SDSS J0836+0054 field is among the densest structures known at these redshifts. One of the spectroscopic companions is a very massive star-forming galaxy ($\mathrm{log}_{10}(\mathcal{M}_{\star}/M_\odot)=10.3_{-0.2}^{+0.3}$) based on its unambiguous detection in a Spitzer 3.6 $\mu$m image. This suggests that the quasar field hosts not one, but at least two rare, massive dark matter halos ($\mathrm{log}_{10}(\mathcal{M}_{h}/M_\odot)\gtrsim12$), corresponding to a galaxy overdensity of at least 20. We discuss the properties of the young radio source. We conclude that the environment of SDSS J0836+0054 resembles, at least qualitatively, the type of conditions that may have spurred the direct collapse of a massive black hole seed according to recent theory. 
\end{abstract}



\section{Introduction} 
\label{sec:intro}

Since their discovery, the study of luminous quasars within the first billion years of cosmic time has intersected with numerous areas in astrophysics from black holes to cosmology \citep{fan00}. Although they are the most luminous hydrogen-ionizing photon producing sources at their  epoch, their relatively low number density impies that they probably did not play a major role in driving the overall reionization of the universe \citep{bouwens15}. Ironically, however, their rest-frame UV spectra offer some of the most direct ways we currently have of assessing the rapidly evolving neutral fraction of the intergalactic medium (IGM) \citep[e.g.,][]{mortlock11,becker15,eilers17,davies18,wang20,yang20,bosman20}. Estimates of the masses of their supermassive black holes (SMBHs) using established techniques indicate that at least some of these objects were able to accumulate masses similar to the SMBH at the center of the local giant elliptical M87 within less than 1 Gyr of the Big Bang, defying the most simple maximum accretion scenarios especially for the highest redshifts \citep[e.g.,][]{banados18}. 

One of the main uncertainties at the current moment relates to the population of ``seed'' black holes from which they originated \citep[e.g.,][]{bromm11,greene12,volonteri12,mezcua17,inayoshi20}. Although in principle a $\sim10^9$ $M_\odot$ SMBH could form from a 100 $M_\odot$ seed accreting at the Eddington rate for under 1 Gyr, it is not clear whether such high accretion rates could be sustained for such a long time (or shorter periods of super-Eddington accretion), especially in the presence of feedback in the forming galaxy \citep{inayoshi16,smith18,regan19}. One possible solution includes mergers between light black hole seeds in clusters of stellar remnants \citep{kroupa20} or at the centers of merging (mini)halos, but long black hole merger time scales and the effect of merger recoil kicks could pose a problem for such a scenario \citep[see][]{piana21}. An attractive, alternative scenario starts with the formation of a much more massive seed of $\sim10^{4-5}$ $M_\odot$, a so-called direct collapse black hole (DCBH) seed \citep[e.g.][]{umemura93,eisenstein95,koushiappas04,volonteri05,lodato06}. In order to accomplish this, models and simulations typically require some kind of mechanism that prevents cooling and fragmentation and stimulates the formation of the DCBH seed, through an enhanced local Lyman-Werner photon flux or through the dynamical heating of merging (mini)halos. These mechanisms, which could operate alone or in tandem, may prevent halos from forming stars at least until reaching the atomic cooling stage, at which point a runaway collapse to a massive seed occurs, either directly or by forming first a shortlived, supermassive star \citep[e.g.,][]{agarwal16,habouzit16,latif18,wise19,lupi21,piana21}.

Based on their low co-moving space density, high accretion power and large black hole masses, it is frequently argued that the first luminous quasars must have formed in very dense regions of the cosmic web \citep[e.g.,][]{fan00,springel05,li07,sijacki09}. It has thus been expected that these luminous quasars will be situated in or near large overdensities of dark matter, gas and galaxies that, in principle, we should be able to detect. Precisely this expectation has been tested for the last two decades in many different and complementary ways, but it has been challenging to interpret that work in a straightforward and clear-cut way \citep[see, e.g.,][and references therein for a variety of observational findings related to the environments of quasars]{overzier09_mr,overzier16,mazzucchelli17,decarli17,mignoli20}.    

These results are even more puzzling in the light of several significant large-scale overdensities that have been detected at similar redshifts with relative ease in fields not known to host luminous quasars \citep[e.g.,][]{ouchi05,toshikawa12,ota18,calvi19}. Although obscuration, variability or duty cycle can easily explain the real or apparent absence of luminous quasars in any of these structures, there are no obvious mechanisms that could temporarily hide any large-scale structure around the known quasars, if present. This is a rather strong, albeit indirect, argument against the explanation that the structures associated with the quasars are simply being missed because we lack observational sensitivity at these redshifts. 

Another common explanation given for the apparent absence of galaxies clustered near quasars is that strong radiative feedback has raised the temperature floor of the surrounding IGM, thereby preventing the condensation of gas into galaxies \citep[e.g.,][]{utsumi10,banados13,mazzucchelli17}. In this case, the quasars would still be surrounded by large matter overdensities as expected from theory, but those would be devoid of galaxies within the quasar ionization cone. In this scenario, however, it seems challenging to suppress any structure on scales out to many Mpc, uniformly around the quasar, and throughout its active and inactive phases as the observations appear to require. Besides, galaxies have been found within the ionization cones of some quasars \citep[e.g.,][]{mignoli20,bosman20}. 

The apparent lack of clearly identifiable overdense environments around most quasars at $z\sim6$ studied to date is even more puzzling given current theories for massive seed formation. As explained above, these theories require large matter overdensities that could (1) provide a large Lyman-Werner photon background, and (2) enhanced merger rates of small dark matter halos that stimulate the collapse of a massive gas cloud into a DCBH seed \citep[e.g.,][]{wise19,regan20,lupi21}. Once the massive seed of $\simeq10^3-10^5$ $M_\odot$ has formed, the high merger and accretion rates in the overdense environment would be sufficient to achieve the 4-6 orders of magnitude growth necessary to form the SMBH in just a few hundred million years. Without invoking the DCBH scenario, it is a real challenge to explain the very large SMBH masses in at least some of the quasars at $z\gtrsim6$. 

Quasar SDSS J0836+0054 at $z=5.8$ (denoted J0836 throughout this paper) is among a handful of objects for which a large potential overdensity of galaxies was found in early observations using the HST \citep[][Z06]{zheng06}. However suggestive this result, direct spectroscopic evidence for an overdensity associated with J0836 was still lacking. Recently, however, three galaxies having redshifts near that of the quasar were found as part of the study of \citet[][B20]{bosman20}, confirming close association with the quasar for two objects from the Z06 sample, plus one new source. As we show in this paper, one of the companion galaxies is very bright at 3.6 $\mu$m, indicative of a very high stellar mass. Based on these new results, we conclude that the reality of a structure of galaxies associated with J0836 has now been established. We compare the structure around J0836 with model expectations and recent literature results, including another recently confirmed large structure associated with quasar SDSS J1030+0524 at $z=6.3$ \citep{mignoli20} as well as several field overdensities at $z\sim6-7$. From these comparisons, we conclude that J0836 lies in a rather exceptional environment that at least qualitatively resembles the type of overdense environments required by the DCBH scenario. 

We set the cosmological parameters to $H_0=69.6$ km s$^{-1}$ Mpc$^{-1}$, $\Omega_M=0.286$, $\Omega_\Lambda=0.714$, which gives a spatial scale of 5.95 physical kpc arcsec$^{-1}$ and an age of 0.98 Gyr at $z=5.8$ \citep{bennett14}. We will use the  prefix `p' to indicate physical scales (e.g., pkpc) and the prefix `c' to indicate co-moving scales (e.g., cMpc). We use AB magnitudes. 

\begin{figure}[t]
\centering
\includegraphics[width=\columnwidth]{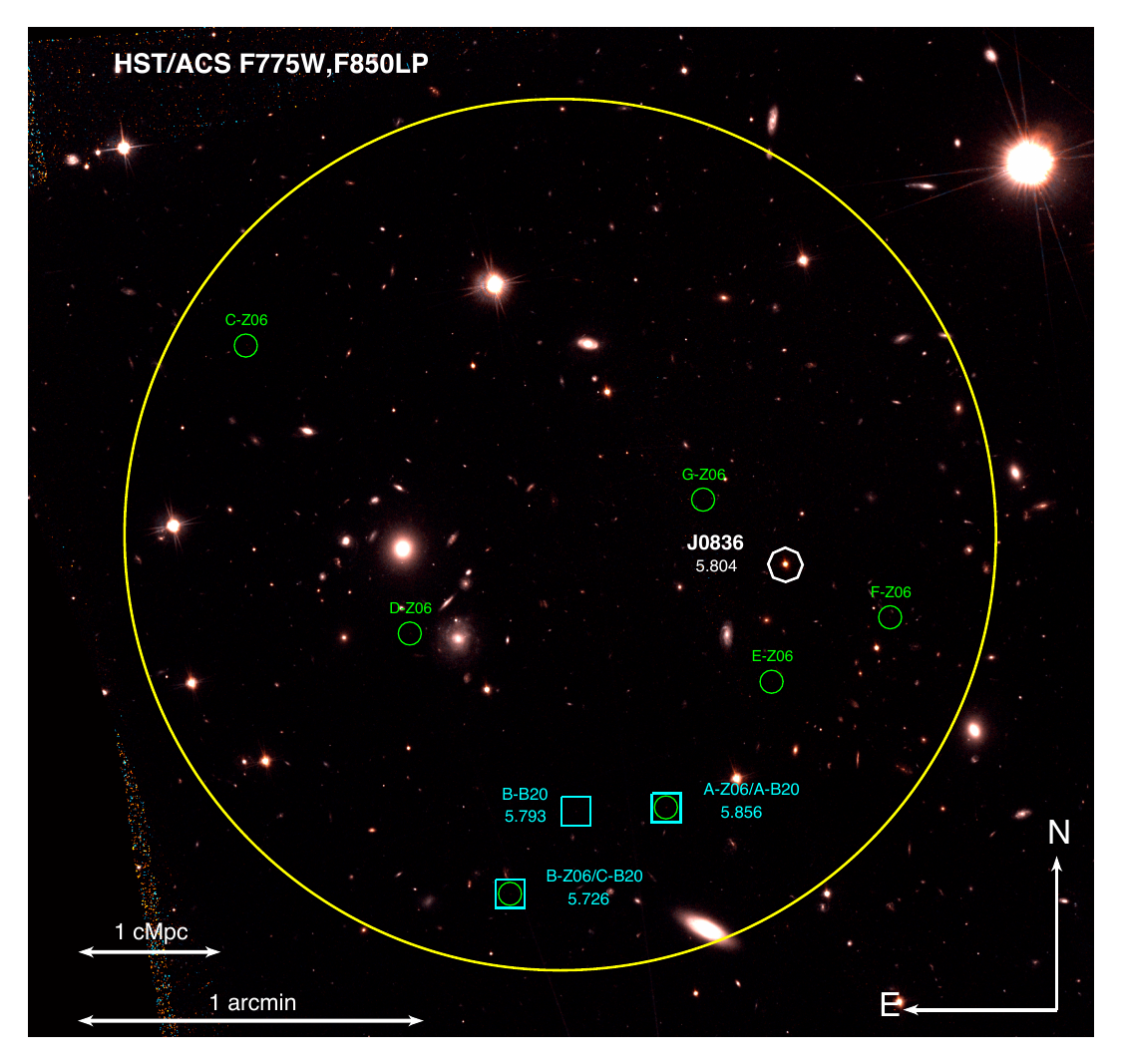}
\caption{HST/ACS false color image of the region around SDSS J0836+0054 at $z=5.8$ using the \ip\ image for the blue channel, (\ip+\zp)/2 as the green channel and \zp\ as the red channel. The quasar is marked by the white polygon. Photometric dropouts from Z06 and spectroscopic \lya\ emitting objects from B20 are indicated by the green circles and cyan squares, respectively. Redshifts from B20 have been indicated. The large yellow circle referred to in this study has a radius of 1\farcm26 (radius of 0.45 pMpc). Scalebars of length 1\arcmin\ and 1 cMpc are shown in the bottom-left.
\label{fig:field}}
\end{figure}

\section{Data and samples}

\subsection{SDSS J083643.85+005453.3 (J0836)}

The quasar J0836 was discovered as part of the Sloan Digital Sky Survey (SDSS) quasar survey \citep{fan01}, and is one of the brightest quasars at $z>5.7$ in the optical \citep{banados16}, radio \citep{banados21} and X-rays \citep{wolf21}. Various redshift determinations in the range $z\simeq5.77-5.83$ exist \citep{fan01,stern03,freudling03,kurk07,jiang07,shen19}, but we follow B20 who determined $z=5.804\pm0.002$ based on the \oii\ and \cii\ emission lines. J0836 is also one of the most distant known radio-loud quasars: it has a flux density of 1.1 mJy at 1.4 GHz, a 5 GHz (rest-frame) luminosity of $1.1\times10^{25}$ W Hz$^{-1}$ sr$^{-1}$, a radio spectral index of -0.9 between 1.5 and 5 GHz, and is marginally resolved at VLBI resolutions of $\sim10$ mas ($\sim$40 pc) \citep{petric03,frey03,frey05}. \citet{wolf21} presented new observations at low radio frequency, showing that the radio spectral index flattens significantly below 1.4 GHz, consistent with a peaked radio spectrum. The compact radio size, steep spectral index and the evidence for a peaked spectrum may suggest that this is a young radio source in which the jets are still confined to (sub-)kpc scales. J0836 has an estimated black hole mass of $\mathrm{log}_{10}(\mathcal{M}_\mathrm{BH}/M_\odot)=9.48\pm0.55$ (see Sect. \ref{sec:quasar} below). 

\begin{table*}[t]
\scriptsize
\centering
\caption{Objects in the field of J0836. \label{tab:sample}} 
\begin{tabular*}{0.9\textwidth}{lccccccc} 
\hline
\hline
ID  & $\alpha$  & $\delta$ &  $z_\mathrm{phot}^a$ & $z_\mathrm{spec}^b$ & $\Delta z^{\dagger\dagger}$ & EW$_{0,\mathrm{Ly}\alpha}^b$ & log($L_{\mathrm{Ly}\alpha}/[\mathrm{erg~s^{-1}}])^b$\\  
    &  (J2000)  & (J2000)  &     &                   &  & (\AA) &   \\ 
\hline   
\hline
J0836 &   08:36:43.871  & +00:54:53.15    &    --   & $5.804\pm0.002$ & 0.0 &--&--\\
\hline
B-B20 & 08:36:46.280  & +00:54:10.55  &       --   & $5.793\pm0.003$ &--0.011 & $76^{+55}_{-34}$ & $43.03_{-0.03}^{+0.03}$\\
A-Z06/A-B20 & 08:36:45.248 & +00:54:10.99  &  $5.8^{+1.4}_{-0.2}$ & $5.856\pm0.003$&+0.052 & $>10.1$  & $42.63^{+0.05}_{-0.04}$\\ 
B-Z06/C-B20 & 08:36:47.053 & +00:53:55.90  &  $5.9^{+1.0}_{-1.0}$ & $5.726\pm0.003$&--0.078 & $55^{+8}_{-5}$ & $42.93^{+0.15}_{-0.11}$\\ 
B$^\prime$-A06  & 08:36:47.127 & +00:53:56.20  &  $5.70^{+0.03}_{-0.05}$ &  -- & -- &--&--\\ 
C-Z06   & 08:36:50.099 & +00:55:31.16  &  $5.9^{+1.1}_{-0.5}$    &  -- & -- &--&--\\ 
C2-Z06  & 08:36:50.058 & +00:55:30.54  &  $5.9^{+1.4}_{-1.5}$    &  -- & -- &--&--\\ 
C3-Z06  & 08:36:50.010 & +00:55:30.27  &  $7.0^{+0.0}_{-0.7}$    &  -- & -- &--&--\\ 
D-Z06   & 08:36:48.211 & +00:54:41.19  &  $5.8^{+1.2}_{-0.7}$    &  -- & -- &--&--\\ 
E-Z06   & 08:36:44.029 & +00:54:32.79  &  $5.2^{+1.7}_{-0.7}$    &  -- & -- &--&--\\ 
F-Z06   & 08:36:42.666 & +00:54:44.00  &  $5.7^{+1.2}_{-0.7}$    &  -- & -- &--&--\\ 
G-Z06$^\dagger$ & 08:36:44.809 & +00:55:04.41&  $5.7^{+1.2}_{-0.8}$    &  -- & -- &--&--\\ 
\hline
\hline
\multicolumn{8}{l}{$^a$ Objects A-Z06 to G-Z06 from \citet{zheng06} and object B$^\prime$-A06 from \citet{ajiki06}.}\\
\multicolumn{8}{l}{$^b$ From \citet{bosman20}.}\\
\multicolumn{8}{l}{$^\dagger$ During this work a mistake was found in the coordinates for object G used in Table 1 and Figure 1 of Z06.}\\
\multicolumn{8}{l}{Here we give the correct coordinates.}\\
\multicolumn{8}{l}{$^{\dagger\dagger}$ Redshift difference of spectroscopic galaxies and J0836, $\Delta z = z_\mathrm{obj}-z_\mathrm{QSO}$.}\\
\end{tabular*}
\end{table*}

\subsection{HST/ACS and Spitzer/IRAC images}

We use images obtained with the HST/ACS in the filters F775W (\ip; 4676 s) and F850LP (\zp; 10,778 s). These data are described in detail in Z06. For the analysis in this paper, we retrieved the pipeline reduced mosaics from the Hubble Legacy Archive\footnote{\url{https://hla.stsci.edu/}}, and performed new measurements using Source Extractor version 2.25.0 \citep{bertin96}. We applied a Galactic extinction correction of 0.1 and 0.07 mag to the \ip\ and \zp\ magnitudes, respectively. We also use Spitzer/IRAC 3.6 $\mu$m observations with a total exposure time of 17 ks taken from \citet{overzier09_irac}. 

\subsection{Galaxy samples and redshifts}

The samples discussed in this paper are the following. We use the sample of photometrically selected \ip-dropouts detected with HST/ACS from Z06. The dropout galaxies have \zp$<$26.5 mag and colors 1.3$<$\ip--\zp$<$2.0, consistent with $z\sim5.8$. There are seven objects in this sample (labeled A--G). Below, when referring to the Z06 objects we will use these IDs together with the suffix ``Z06". One of the dropouts (C-Z06) has two close companions in projection (labeled C2 and C3) that are fainter than the imposed limit in \zp.

\citet{ajiki06} carried out broad and narrowband observations using the Subaru Suprime-Cam with filters $B$,$V$,$r^\prime$,$i^\prime$,$z^\prime$, and the narrowband filter $NB816$ sensitive to \lya\ at $z=5.65-5.75$. They found one strong \lya\ emitting candidate at $z\approx5.7$ about 1\arcsec\ or 6 kpc from object B-Z06. We will refer to this object as B$^\prime$-A06. The remainder of the Z06 objects were not detected in $B$,$V$,$r^\prime$, which is consistent with them being genuine $i$-dropouts and not foreground interlopers. 

Finally, and key to the results presented in this paper, we use three spectroscopically confirmed \lya\ emitting objects from B20. Their selection was also based on the Subaru data from \citet{ajiki06}. They used color selection criteria optimized to the redshift range $5.65\le z\le 5.90$, and supplemented spectroscopic targets with a photometric redshift selection. Of 19 candidates found in the region overlapping with the HST data, 3 corresponded to objects from Z06: A-Z06, B-Z06 and F-Z06. 11 of these were targeted with the Keck/DEIMOS, including A-Z06 and B-Z06. When referring to objects from the spectroscopic sample, we will use the suffix ``B20". Redshifts were obtained for three objects: ``Aerith A" at $z=5.856\pm0.003$ (A-B20 which is identical to A-Z06), ``Aerith B" at $z=5.793\pm0.003$ (B-B20 has no counterpart in Z06 because of its relatively small \ip--\zp\ color) and ``Aerith C" at $z=5.726\pm0.003$ (C-B20 which is identical to B-Z06). 

Object A-Z06/A-B20 is also detected in the NB816 narrowband, which is interpreted as continuum emission blueward of \lya\ \citep{bosman20}. Object B-B20 is not detected in the NB816. The redshift of B-Z06/C-B20 places the wavelength of its \lya\ emission in the NB816. A relatively bright object (B$^\prime$-A06) has indeed been detected, as first reported by \citet{ajiki06}. However, as explained in B20, the redshift of B-Z06/C-B20 was obtained at the location of its $z^\prime$-band continuum, which is offset from the NB816 source B$^\prime$-A06 by about 1\arcsec. It is thus possible that there are two galaxies at this location separated by about 6.3 kpc if at the same redshift. B20 report no other objects with redshifts close to that of J0836, while \citet{meyer20} report on a fourth source at $z=5.284$ that is not relevant to this paper.

Details on the samples are summarized in Table \ref{tab:sample}.

\section{Analysis and results}
\label{sec:3}

\subsection{Basic structure}

\begin{figure}[t]
\centering
\includegraphics[width=\columnwidth]{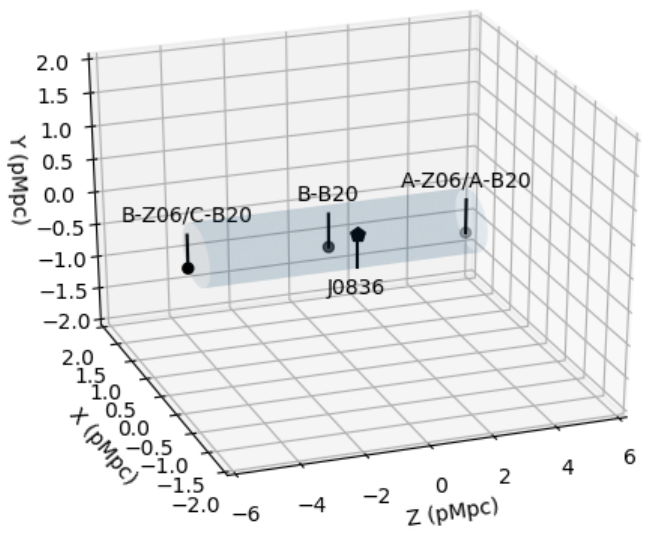}
\caption{Locations of the spectroscopically identified galaxies with respect to the quasar in proper physical coordinates. The $Z$-axis was chosen for the direction along the line of sight, and the $X$- and $Y$-axes were chosen to indicate the directions along Right Ascension and Declination, respectively. The structure can be approximated by a narrow cylinder elongated along the line of sight. When this cylinder is seen face-on, as in the case of the HST observations shown in Fig. \ref{fig:field}, the structure appears highly compact.     
\label{fig:proj}}
\end{figure}

In Fig. \ref{fig:field} we show a false color image of the archival HST/ACS \ip\ and \zp\ bands of the J0836 field. The quasar itself is marked by a white polygon. The \ip-dropouts from Z06 are marked by green circles, and the spectroscopically confirmed objects from B20 by cyan squares. The work of B20 is significant because it has confirmed, for the first time, that 2 out of the 7 objects from Z06 have redshifts close to that of the quasar (much closer than one would expect for a randomly distributed sample of \ip-dropouts, see below). Moreover, they also found an additional object (B-B20), that lies closer to the quasar redshift than the other two. The proper distances along and perpendicular to the line of sight (l.o.s.) to the quasar ($R_\parallel$, $R_\perp$) are (3.64,0.28) pMpc for A-Z06/A-B20, (--0.67,0.33) pMpc for B-B20 and (--5.03,0.44) pMpc for B-Z06/C-B20, where a minus sign indicates objects in the foreground of the quasar. As further illustrated by Fig. \ref{fig:proj}, the three objects lie along a narrow cylinder (radius of 0.5 pMpc) oriented along the line of sight with a total length of about 8.7 pMpc (corresponding to the maximum redshift difference of $\Delta z=0.13$) and the quasar about mid-way. The length of this cylinder is on the larger side of the distribution of cosmic filament sizes found in cosmological simulations \citep{galarraga20}. However, we will refrain from calling this structure a filament because of the relatively small number of redshifts it is based on. Future data may show that the true three-dimensional structure around J0836 is very different from the simple `filament' drawn in Fig. \ref{fig:proj}.  

\subsection{The quasar}
\label{sec:quasar}

\citet{kurk07} estimated a black hole mass of $\mathrm{log}_{10}(\mathcal{M}_\mathrm{BH}/M_\odot)=9.4\pm0.1$ based on the \mgii\ line. We have updated their estimate using newer measurements and calibrations. We use the black hole mass calibration from \citet{vestergaard09} together with the FWHM(\mgii)$=3600\pm300$ measurement of \citet{kurk07} and the 3000 \AA\ rest-frame luminosity of $\mathrm{log}_{10}({L}/\mathrm{[erg~s^{-1}]})=47.0$ from \citet{shen19}, finding $\mathrm{log}_{10}(\mathcal{M}_\mathrm{BH}/M_\odot)=9.48\pm0.55$, consistent with the estimate of \citet{kurk07}, but $\sim$0.4 dex below the estimate of \citet{shen19} based on the problematic \civ\ line.

A dynamical mass estimate for the quasar host galaxy does not yet exist, but an order of magnitude estimate can be made by taking the results for a sample of $z\sim6$ quasars from \citet{neeleman21} \citep[see also][]{venemans16}. They find that, on average, the quasar host galaxy dynamical masses are a factor of 7 lower than expected for their black hole mass and assuming the \citet{kormendy13} $\mathcal{M}_\mathrm{BH}-\mathcal{M}_\star$ relation for local bulges. With an estimated black hole mass of $\mathrm{log}_{10}(\mathcal{M}_\mathrm{BH}/M_\odot)\sim9.5$, this would imply a stellar (bulge) mass of $\mathrm{log}_{10}(\mathcal{M}_\star/M_\odot)\sim10.8$. If we assume a typical stellar-to-dark-matter mass ratio of $\sim30-50\times10^{-3}$ appropriate for the high mass end of the halo mass function at $z\sim6$ \citep{stefanon21}, this would  imply a quasar host halo mass of $\mathrm{log}_{10}(\mathcal{M}_h/M_\odot)=12-12.3$. Although the uncertainty of this estimate is large, it is consistent with recent derivations based on \ciialma\ velocity widths from \citet{shimasaku19} and proximity zone measurements from \citet{chen21}.

\begin{figure}[t]
\centering
\includegraphics[width=0.8\columnwidth]{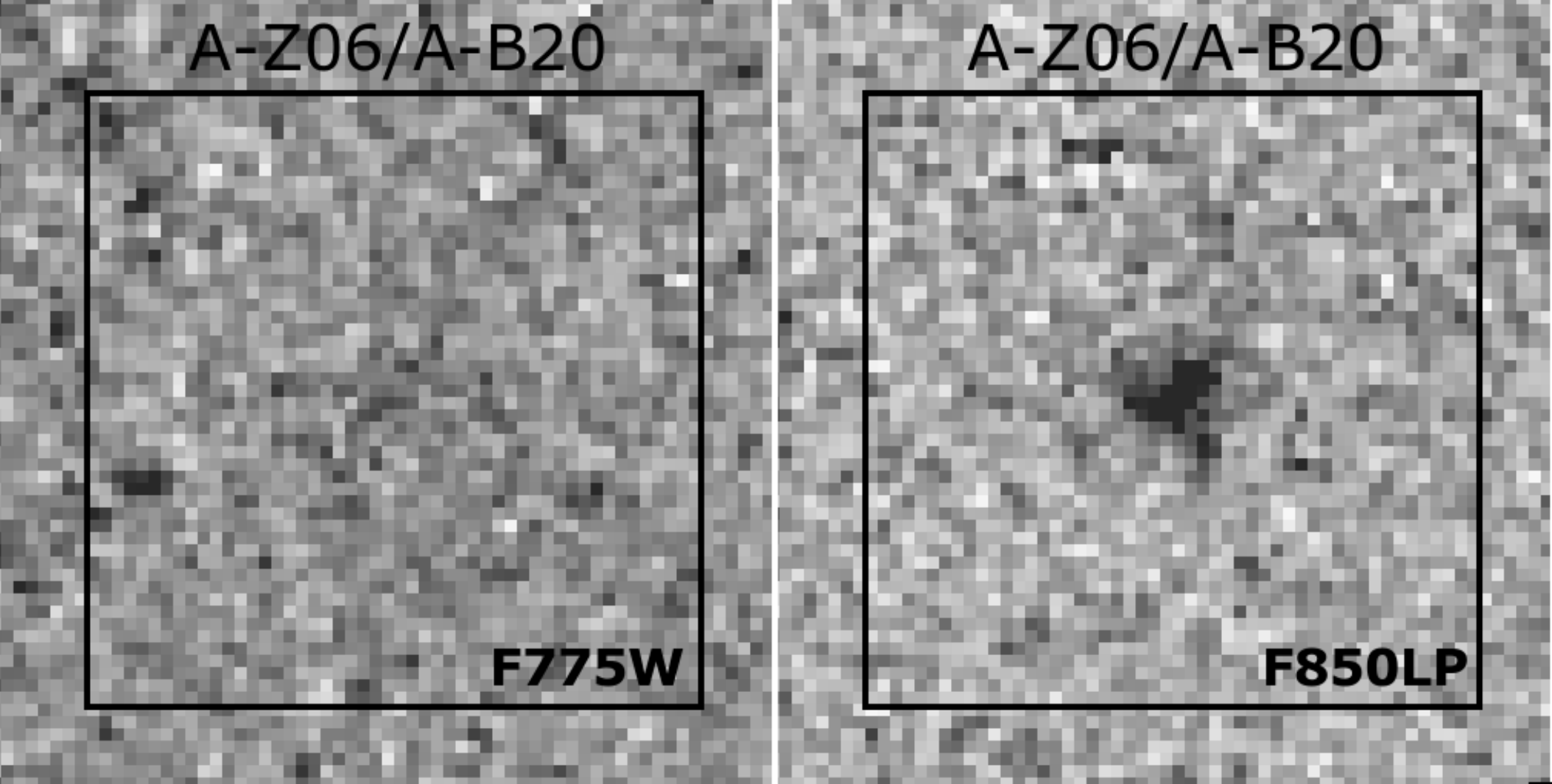}\\
\vspace{2mm}
\includegraphics[width=0.8\columnwidth]{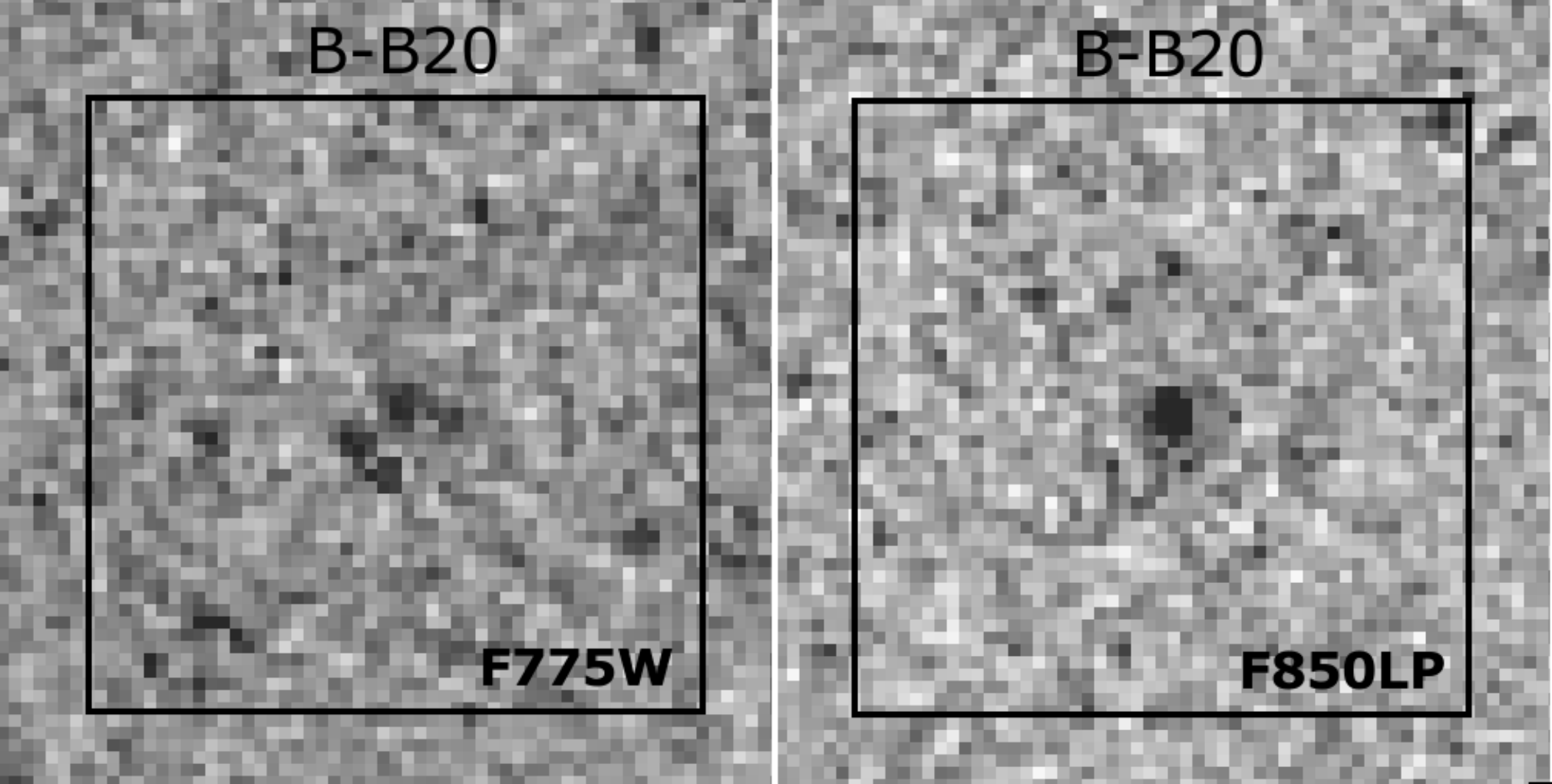}\\
\vspace{2mm}
\includegraphics[width=0.8\columnwidth]{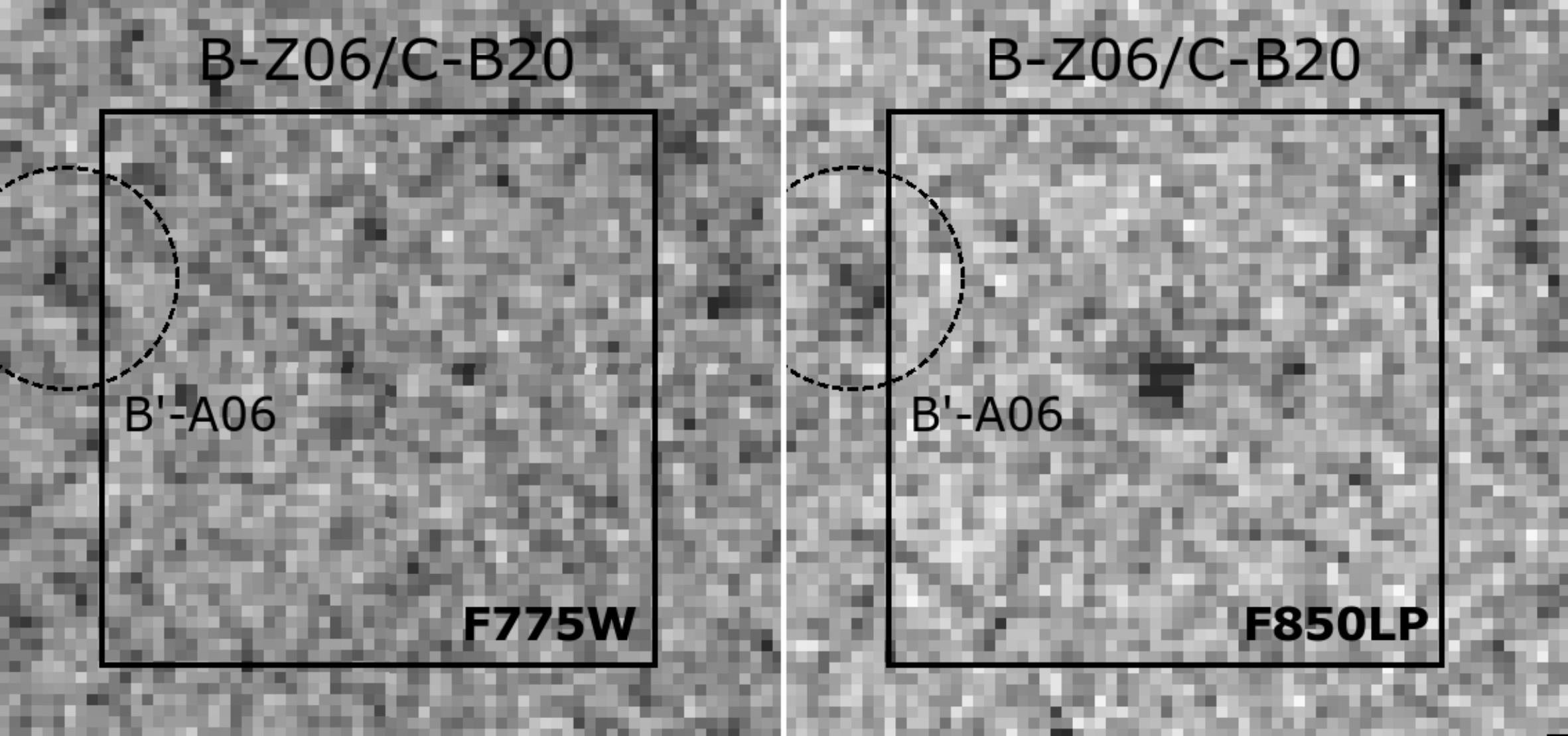}
\caption{HST/ACS image stamps in \ip\ (left) and \zp\ (right) of the three sources confirmed by B20 with A-Z06/A-B20 in the top, B-B20 (no Z06 counterpart) in the middle, and B-Z06/C-B20 in the bottom panels. The boxes measure 2\arcsec$\times$2\arcsec. The image pixel size is 0\farcs04 and the data is shown unsmoothed. In the bottom panels, the dashed circle marks the narrow-band excess object B$^\prime$-A06 about 1\arcsec\ East of B-Z06/C-B20 detected in the Subaru NB816 image by \citet{ajiki06}. See text for details. \label{fig:stamps}}
\end{figure}

\begin{table}[t]
\begin{center}
\caption{Magnitudes and sizes of the confirmed objects.\label{tab:stamps}} 
\begin{tabular}{lcccc} 
\hline
\hline
ID & Filter & $r_{50}^\dagger$ & $r_{90}^\dagger$ & $m_{AB}$ \\ 
   &        & (arcsec) & (arcsec) & (mag) \\
\hline
\hline
A-Z06/A-B20  & \ip & 0.20  & 0.37  & $26.84\pm0.17$ \\ 
       & \zp & 0.16  & 0.38  & $25.44\pm0.05$ \\ 
       & 3.6 $\mu$m & -- & -- & $23.78\pm0.09$\\
\hline
B-B20  & \ip & 0.16  & 0.26  & $26.57\pm0.12$ \\ 
       & \zp & 0.12  & 0.29  & $26.07\pm0.08$ \\ 
\hline
B-Z06/C-B20  & \ip & 0.13  & 0.39  & $27.58\pm0.36$ \\ 
       & \zp & 0.18  & 0.38  & $25.89\pm0.08$ \\ 
B$^\prime$-A06 & \ip & 0.10  & 0.35  & $27.53\pm0.20$ \\ 
       & \zp & 0.10  & 0.16  & $27.14\pm0.15$ \\ 
\hline
\hline
\end{tabular}
\end{center}
$^\dagger$ The sizes quoted are the radii containing 50 and 90\% of the total flux as returned by Source Extractor. They were not corrected for the HST seeing of about $0\farcs07$ (FWHM).  
\end{table}

\subsection{Massive companion galaxy A}
\label{sec:obja}

\begin{figure}[t]
\centering
\includegraphics[width=\columnwidth]{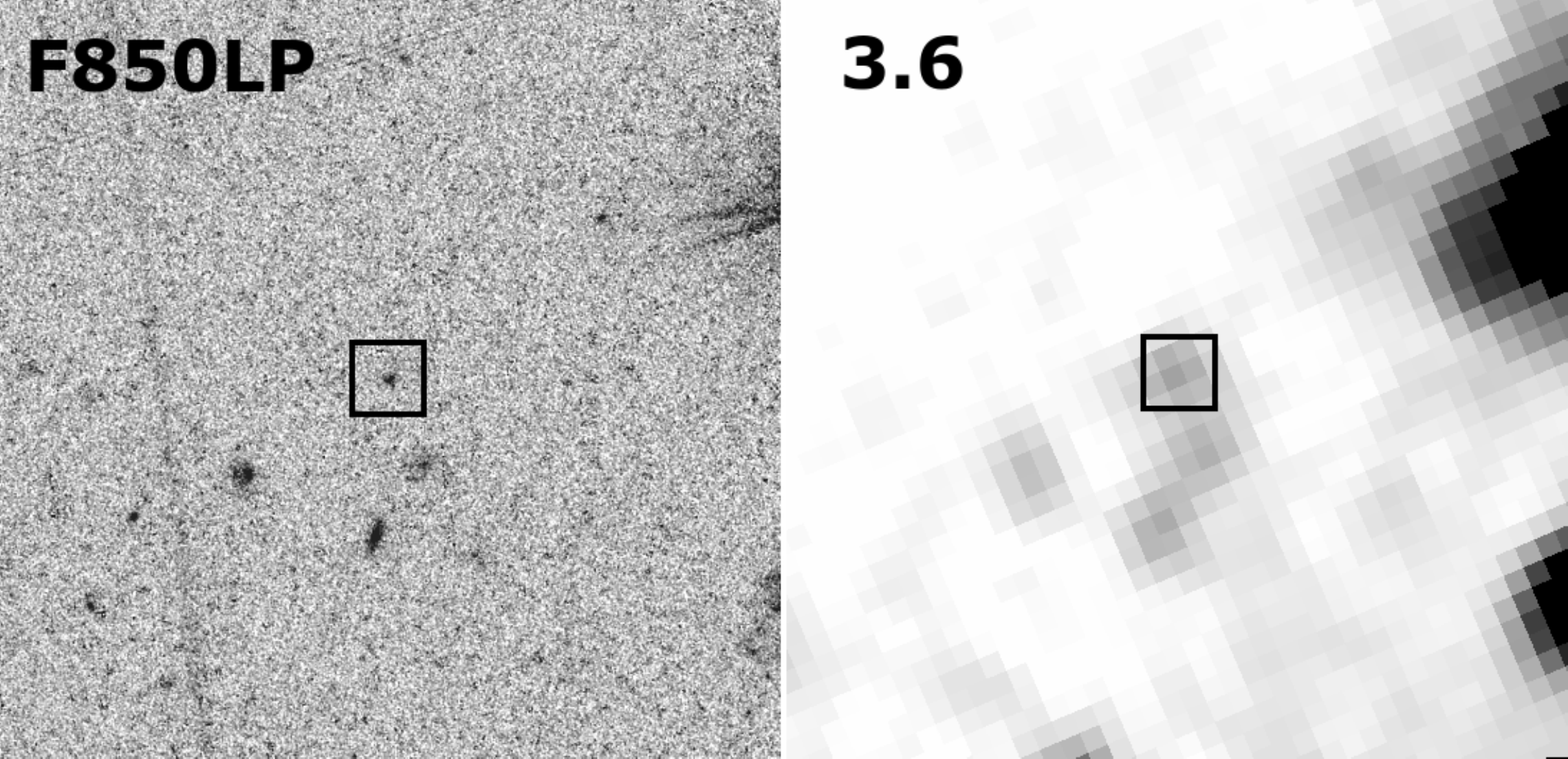}
\caption{HST/ACS \zp\ and Spitzer/IRAC 3.6 $\mu$m images of the object A-Z06/A-B20 discussed in Sect. \ref{sec:obja}. The boxes measure 2\arcsec$\times$2\arcsec\ as in Fig. \ref{fig:stamps}. The HST object is clearly detected at 3.6 $\mu$m. The (deblended) object flux was extracted by simultaneously fitting point sources at the locations of the four objects seen near the central part of the image \citep[see][]{overzier09_irac}.\label{fig:irac}}
\end{figure}

\citet{overzier09_irac} presented Spitzer/IRAC observations of the J0836 structure, and showed that object A-Z06/A-B20 is very bright at 3.6 $\mu$m with $m_{3.6}=23.78\pm0.09$ mag. Assuming that its redshift was indeed $z\sim6$, they concluded that it is among the brightest and most massive \ip-dropout galaxies. Here we will revisit the analysis of A-Z06/A-B20. The \zp\ and 3.6 $\mu$m images are shown in Fig. \ref{fig:irac}. 

\begin{figure}[t]
\centering
\includegraphics[width=\columnwidth]{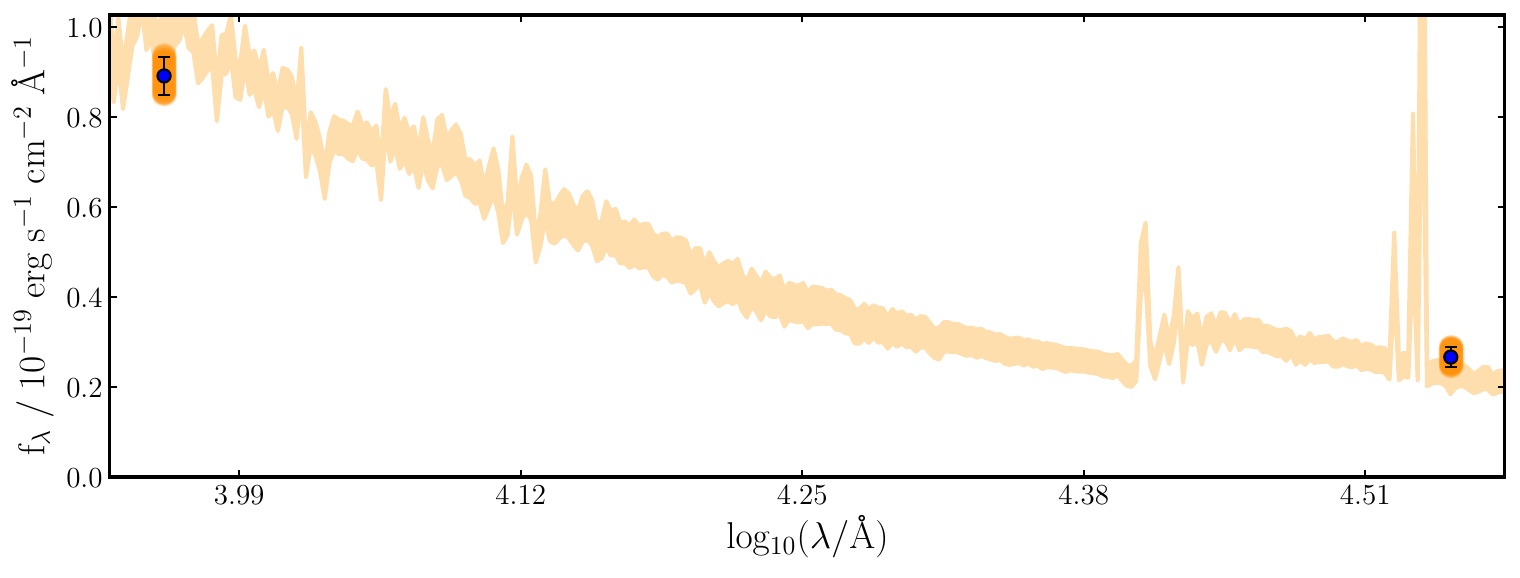}
\includegraphics[width=\columnwidth]{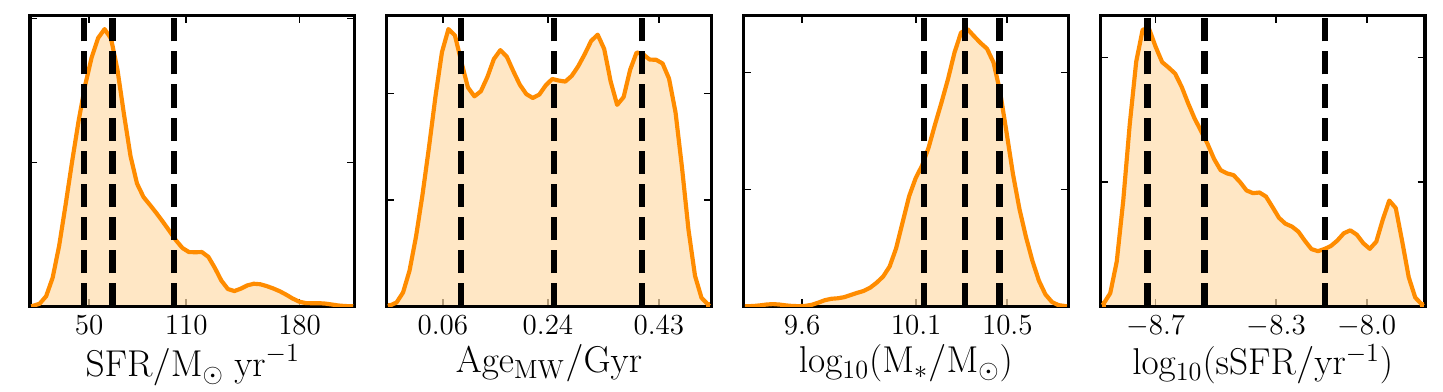}
\caption{Results of the constant star formation model fits to the \zp\ and 3.6 $\mu$m fluxes of object A-Z06/A-B20 at $z=5.856$ described in Sect. \ref{sec:obja}. The top panel shows the best-fit spectral energy distribution (curve) together with the photometry and the sampled posterior probability distribution (data points with shaded regions). The bottom panels show the probability distributions of the SFR, mass-weighted age, stellar mass and specific SFR with the 16, 50 and 84 percentiles indicated by dashed vertical lines.\label{fig:afit}}
\end{figure}

Using the redshift firmly established by B20, we can now make a much more secure determination of the properties of A-Z06/A-B20 based on the available photometry. Although the \zp--[3.6] color is sensitive to age, dust, SFH and metallicity (and redshift), the 3.6 $\mu$m flux probing rest-frames of around 5000 \AA\ is an accurate (albeit relative) gauge of stellar mass at these redshifts \citep[e.g., see][]{overzier09_irac,duncan14,bhatawdekar19,stefanon21}. We used the BagPipes code \citep{carnall18} to fit the \zp\ and 3.6 $\mu$m photometry. The \ip-band was excluded because it does not add any information about the intrinsic SED. The redshift was fixed to $z=5.856$. We fit constant and exponential star formation histories with a \citet{kroupa02} IMF. We set the maximum age to the age of the universe at $z=5.8$. Because galaxies at $z\sim6$ are known to have a relatively strong contribution from emission lines (mainly \oiii) to their 3.6 $\mu$m broadband flux, we used the option to include nebular emission with an ionization parameter $U=0.01$. Assuming a constant star formation history and attenuation by dust according to the \citet{calzetti00} law, we find a stellar mass $\mathrm{log}_{10}(\mathcal{M}_\star/M_\odot)=10.34_{-0.16}^{+0.30}$ with a mass-weighted age of $0.27_{-0.14}^{+0.28}$ Gyr ($z_f\approx7.5$), a metallicity $\mathrm{log}_{10}(Z/Z_\odot)=0.86_{-0.45}^{+0.86}$, and $A_V=0.7_{-0.1}^{+0.3}$ mag (see Fig. \ref{fig:afit}). The stellar mass and mass-weighted age returned for an exponential star formation history model are essentially the same. Without the nebular emission, the stellar mass is about 0.1 dex higher. Restricting the metallicity to $\le0.4Z_\odot$ as in \citet{stefanon21}, we find a stellar mass $\mathrm{log}_{10}(\mathcal{M}_\star/M_\odot)=10.2_{-0.2}^{+0.3}$ ($\mathrm{log}(\mathcal{M}_\star/M_\odot)=10.40_{-0.14}^{+0.23}$) with (without) the nebular emission with little change to the other parameters. Below we will use our first estimate of $\mathrm{log}_{10}(\mathcal{M}_\star/M_\odot)=10.34_{-0.16}^{+0.30}$ for the stellar mass of A-Z06/A-B20, as the other estimates are consistent within the errors. This stellar mass is two orders of magnitude higher than the minimum halo mass that was inferred by B20.

We can compare these results with recent results for the stellar mass function and stellar-to-halo mass ratios of galaxies at $z\sim5.8$ in the field from \citet{stefanon21}. The stellar mass of A-Z06/A-B20 places it in the highest mass bin considered in that paper ($\mathrm{log}_{10}(\mathcal{M}_\star/M_\odot)=10.4\pm0.2)$, taking into account a division by a factor of 1.5 to convert from their Salpeter to our kroupa IMF. If we use their best-fit $\mathrm{log}(\mathcal{M_\star}/M_\odot)-M_{UV}$ relation with $M_{UV}=-21.3$ mag measured for A-Z06/A-B20, we would naively expect $\mathrm{log}_{10}(\mathcal{M}_\star/M_\odot)\approx9.3$, a full 1 dex below the actual mass measurement and thus indicative of the significant obscuration in this source. 

The stellar mass estimate allows us to make an estimate of the mass of the typical dark matter halos that host objects like A-Z06/A-B20. Using an abundance-matching method, \citet{stefanon21} provide an estimate of the typical stellar-to-halo mass ratio for objects of this mass of $\sim30-50\times10^{-3}$, implying that A-Z06/A-B20 is hosted by a dark matter halo of $\mathrm{log}_{10}(\mathcal{M}_h/M_\odot)\sim12$. The J0836 field thus hosts at least one other massive dark matter halo, besides the quasar itself. Halos of this mass correspond to a virial radius of $r_{200}\simeq50$ kpc, so the halos are really two individual halos given that their separation is about 3.6 pMpc. It is worth pointing out that according to theoretical predictions \citep{piana21}, the black hole occupation fraction of halos of this mass is around 1. It is thus possible that A-Z06/A-B20 also hosts a SMBH, but there is no evidence that it is currently active. 

Although there are no structural measurements for the quasar host galaxy, for object A-Z06/A-B20 we measured a seeing-corrected half-light radius of 0.14\arcsec\ in the \zp-band. This corresponds to a physical half-light radius of 0.83 kpc, giving a stellar mass surface density of $\Sigma_\star\lesssim\mathcal{M}_\star/(2\pi r_{50}^2)=10^{9.8}$ $M_\odot$ kpc$^{-2}$. This should be considered an upper limit because the optical size is likely larger than that measured in the UV. These measurements place object A-Z06/A-B20 near the transition region between massive (compact) quiescent and star-forming galaxies at $z\sim4$ \citep{straatman15}, indicating that A-Z06/A-B20 may be near the end of its active star-forming phase.   

\subsection{Other companions}

We checked the IRAC image for the other confirmed companions, but object B-B20 was not detected and object B-Z06/C-B20 (and thus also B$^\prime$-A06) lies too close to a very bright foreground object to attempt a flux measurement. A direct stellar mass estimate can thus not be made. However, assuming that, unlike object A-Z06/A-B20, these objects do follow the $\mathrm{log}_{10}(\mathcal{M_\star}/M_\odot)-M_{UV}$ relation of \citet{stefanon21} (reddening by dust can be neglected at these magnitudes, \citet{bouwens07}), the stellar masses would be a few times $10^9$ $M_\odot$ for B-B20 and B-Z06/C-B20 and a few times $10^8$ $M_\odot$ for B$^\prime$-A06. The halo masses would be a few times $10^{11}$ and $10^{10}$ $M_\odot$, respectively, based on the abundance matching results from \citet{stefanon21}. The HST images of the objects are shown in Fig. \ref{fig:stamps} and basic measurements are given in Table \ref{tab:stamps}. 

In Sect. \ref{sec:density} below we will analyze the J0836 structure in terms of the probabilities of finding objects like the massive companion A-Z06/A-B20 and the more typical star-forming galaxies B-Z06/C-B20 and B-B20. 

\section{Overdensity analysis}
\label{sec:density}

Several past studies have evaluated the environment of J0836. Z06 found 7 photometrically-selected \ip-dropouts brighter than \zp$\sim$26.5 mag in the central 5 arcmin$^2$ region around J0836 (see Fig. \ref{fig:field}). These 7 objects represent a factor 6 overdensity based on a comparison with the much larger HST Great Observatories Origins Deep Survey \citep[GOODS;][]{giavalisco04} that was observed to a similar depth over an area of 316 arcmin$^2$. 
Using a large \ip-dropout sample extracted from GOODS by \citet{bouwens06}, Z06 found that no single 5 arcmin$^2$ region (a circle with a radius of 1\farcm26) randomly drawn from GOODS contained as many as the 7 objects encountered in the J0836 field, with 4 being the highest. Furthermore, \citet{overzier09_mr} constructed a 4\fdg4$\times$4\fdg4 mock survey ($\approx220\times$\ the area observed by GOODS) of galaxies at $z\sim6$ and showed that (surface) overdensities as found near J0836 are indeed very rare. However, in this large simulation even rarer regions were found due to extremely dense regions mostly associated with (forming) galaxy clusters. However, a definitive conclusion about the nature of the J0836 overdensity was not possible due to the lack of spectroscopic redshifts.  

With the new spectroscopic data from B20, we can now quantify the nature of the overdensity associated with J0836 much more precisely. It is interesting that while only 2 of the 7 objects from Z06 were targeted spectroscopically by B20, they represent 2/3 of the spectroscopically confirmed objects near J0836. These two objects are also the brightest among the Z06 sample (\zp-band magnitudes of 25.5 and 26 mag). The typical surface density and redshift distribution of \ip-dropouts selected with HST is well known. According to \citet{bouwens07}, the cumulative surface density is $0.328\pm0.044$ arcmin$^{-2}$ to \zp=26.5 mag, and based on their redshift distribution the (random) probability of finding a dropout in the redshift range $z=5.73-5.86$ ($\Delta z=0.13$) is 0.16. In this redshift range and in an area of 5 arcmin$^2$ we would thus naively expect to find $0.3\pm0.04$ objects. With two Z06 objects spectroscopically confirmed in this region, the three-dimensional overdensity thus appears sigificant as well ($\delta=n/\bar{n}-1\approx6$). Here we did not take into account the third confirmed object from B20 because it did not pass the \ip-dropout selection criteria used in Z06.   

In the analysis below, the terms dropout and Lyman Break Galaxy (LBG) are used interchangeably to indicate continuum-selected star-forming galaxies. We will use the term \lya\ emitter (LAE) for the subset of star-forming galaxies with a prominent \lya\ emission line having a minimum rest-frame equivalent width (EW) of $\sim$20 \AA. Using this criterion, at least 2 of the 3 B20 objects would classify as LAEs. We assume the redshift distribution of \citet{bouwens07} for \ip-dropouts and that of \citet{inami17} as presented by \citet{mignoli20} for LAEs. According to these distributions the probabilities of finding a single LBG and LAE in the redshift range $z=5.73-5.86$ are 0.16 and 0.073, respectively. Of the 7 Z06 LBGs, two spectra were taken and both objects fell into this range, giving a probability of 0.026 based on the binomial distribution (2 out of 2 with $P=0.16$). Among the spectroscopic targets of B20, four objects with \lya\ were found, three of which fell in the redshift range $\Delta z=0.13$ near J0836. The probability of this occurrence at random is 0.0015 again using binomial statistics (3 out of 4 with $P=0.073$). 

These small (but non-zero) probabilities are suggestive that the presence of the quasar J0836 has some effect on the clustering observed. 

\subsection{Clustering of two massive halos}

In Section \ref{sec:obja} we showed that one of the J0836 companions is an extremely massive galaxy, and the close clustering between the quasar and this object can provide  additional strong constraints on the overdensity in this region. Above we showed that the massive companion A-Z06/A-B20 has a stellar mass of around $\mathrm{log}_{10}(\mathcal{M}_\star/M_\odot)=10.34_{-0.16}^{+0.30}$ and inferred halo mass $\mathrm{log}_{10}(\mathcal{M}_h/M_\odot)\sim10^{12}$. Based on the analysis of several hundreds of square arcminutes with deep HST and Spitzer/IRAC coverage, \citet{stefanon21} show how rare such massive objects are: the number density of \ip-dropouts with $\mathrm{log}_{10}(\mathcal{M}_\star/M_\odot)=10.4\pm0.2$ is $0.06_{-0.05}^{+0.14}\times10^{-4}$ dex$^{-1}$ Mpc$^{-3}$. The volume of the J0836 structure is approximately 1724 cMpc$^3$, estimated by taking a cylinder with radius of 3.1 cMpc and a length of 58.5 cMpc (corresponding to $\Delta z=0.13$). In this volume we thus expect to find $0.002-0.035$ of such massive \ip-dropouts at random, while at least 1 was found (object A-Z06/A-B20) not counting the quasar. The discovery of such a rare massive object as part of the J0836 structure is thus additional evidence that the environment of the quasar is exceptional. 

We can try to estimate how exceptional the J0836 environment is based on the clustering statistics of quasars and galaxies. The typical overdensity of galaxies found in a biased region depends on the amplitude of the quasar-galaxy cross-correlation function, $\xi^{QG}(r)=(r/r_0^{QG})^{-\gamma}$, and the volume $V_\mathrm{eff}$:

\begin{eqnarray}
\langle\delta_g\rangle & = & \frac{1}{V_\mathrm{eff}} \int_V \xi^{QG}(r) dV \nonumber\\
 & = & \frac{1}{V_\mathrm{eff}} \int_0^R 2\pi R \mathrm{d}R \int_0^Z \mathrm{d}Z \left(\frac{\sqrt{R^2+Z^2}}{r_0^{QG}}\right)^{-\gamma}
 \label{eq:density}
\end{eqnarray}

The quasar-galaxy cross-correlation length has not been measured directly for the redshift and samples of our interest, but assuming that the clustering of both the quasars and the galaxies is described by power laws with similar slopes, their cross-correlation length is given by $r_0^{QG}=\sqrt{r_0^{GG}r_0^{QQ}}$. Using the results from galaxy clustering measurements together with a halo occupation distribution model, \citet{harikane21} find that dark matter halos with average masses in the range $\mathrm{log}_{10}(\mathcal{M}_h/M_\odot)\approx11.8-12.2$ at $z\approx5.9$ correspond to bias parameters of 7.7--10.1. A similar result is obtained using the Colossus\footnote{\url{http://bitbucket.org/bdiemer/colossus}} cosmological framework \citep{diemer18}, which gives a bias of 8.7 for halos of $\mathrm{log}_{10}(\mathcal{M}_h/M_\odot)=12$ (peak-height of 5.5). The auto-correlation length of $z\sim6$ galaxies with a similar bias parameter or halo mass was measured by \citet{khostovan19}. They found $r_0=15.56_{-2.71}^{+2.51}$ $h^{-1}$ Mpc ($r_0=16.16_{-3.52}^{+3.80}$ $h^{-1}$ Mpc) for LAEs with halo mass $\mathrm{log}_{10}(\mathcal{M}_h/M_\odot)\approx12.3\pm0.2$ ($\mathrm{log}_{10}(\mathcal{M}_h/M_\odot)\approx12.4\pm0.3$) at $z\approx5.79$ ($z\approx5.56$). In the analysis below we will therefore take $r_0^{GG}\approx15$ $h^{-1}$ cMpc as an approximate value. With a quasar auto-correlation length of $r_0^{QQ}\approx22.3$ h$^{-1}$ cMpc \citep[see][]{garcia-vergara17}, the quasar-galaxy cross-correlation length in Eq. \ref{eq:density} becomes $r_0^{QG}\approx26.1$ cMpc. If we assume a clustering power law slope $\gamma=1.8$, and a cylindrical volume with radius $R=3.1$ cMpc and length $Z=58.5$ cMpc, 
the overdensity of $10^{12}$ $M_\odot$ halos expected in the J0836 field (not counting the quasar) is $\langle\delta_g\rangle = 23.8$ (29.6 for $\gamma=2.0$). 

From this we can then estimate the absolute number of such galaxies expected:

\begin{equation}
N_g = (\langle\delta\rangle+1)\bar{n}_gV_\mathrm{eff}
\label{eq:ng}
\end{equation}

For an average field density of $\bar{n}_g=1.1-3.7\times10^{-5}$ cMpc$^{-3}$ \citep[appropriate for objects of halo mass of order $10^{12}$ $M_\odot$;][]{khostovan19,stefanon21}, we then expect to find 0.5--1.6 of such objects in our cylindrical volume around the quasar, matching the one object that was found. 

We can do a similar exercise for star-forming galaxies in more typical halos. The average co-moving abundance of dropouts with $M_{UV}<-20.13$ is $\bar{n}_g\approx4\times10^{-4}$ cMpc$^{-3}$. Assuming a typical bias of 5 and a correlation length of 5.5 $h^{-1}$ cMpc, appropriate for this population of relatively bright dropout galaxies in halos of around $10^{11}$ $M_\odot$ at $z\sim6$ \citep{overzier06b,barone14,khostovan19}, the quasar-galaxy cross-correlation length in Eq. \ref{eq:density} would be 15.8 cMpc. 

\begin{figure}[t]
\includegraphics[width=\columnwidth]{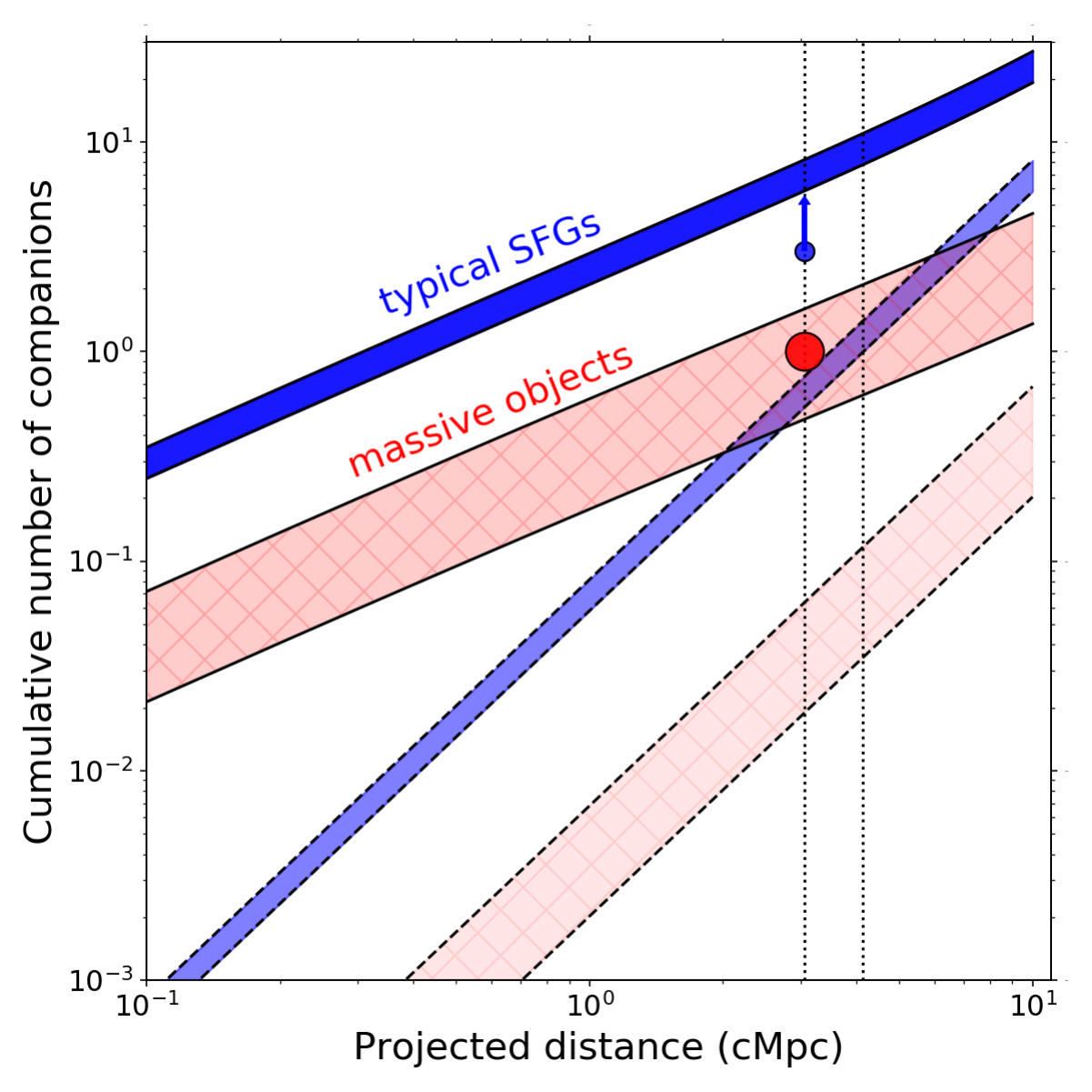}
\caption{Number of companion galaxies expected in a cylinder of length 58.5 cMpc as a function of its projected radius. The blue shaded regions correspond to typical galaxies in the case of no clustering (bottom) and in the case of strong cross-clustering between quasars and galaxies (top). The red hatched shaded regions correspond to massive companion galaxies in the case of no clustering (bottom) and in the case of strong cross-clustering between quasars and galaxies (top). Vertical dotted lines mark radii of 3.1 and 4.1 cMpc, corresponding to the fiducial cylinder used for the J0836 structure in this paper and the half-width of the HST/ACS field, respectively. The large red circle represents the massive companion object A-Z06/A-B20. The small blue circle represents the three spectroscopically confirmed objects from B20. The widths of the shaded regions reflect the uncertainties in the average number densities, and not the uncertainties in the correlation amplitudes. See Sect. \ref{sec:obja} for details. This figure was inspired by a similar figure in \citet{decarli17}.\label{fig:density}}
\end{figure}

In Fig. \ref{fig:density} we show the cumulative number of massive and typical companion galaxies expected in the J0836 field following these calculations with the red- and blue-shaded regions, respectively. We plot the number of objects as a function of the projected co-moving radius of a cylinder with a fixed length of 58.5 cMpc. The finding of 1 massive companion within 3.1 cMpc is consistent with the enhancement in clustering over the random expectation due to the presence of the massive quasar halo (upper red shaded curve and large circle). Doing the same calculation for more typical star-forming galaxies in lower mass halos, the number of 3 objects detected (small blue circle with arrow) is lower than that expected (upper blue-shaded region) by a factor of 2--3, but it is important to keep in mind that the spectroscopic sample is highly incomplete (for example, there are at least an additional 5 objects from Z06 that have not been targeted). 

This analysis shows that the clustering observed in the J0836 field is consistent with what we would expect from the presence of a massive quasar halo. In Section \ref{sec:comparison} below we will compare the J0836 structure to several other dense structures of galaxies discovered in recent years, both near quasars and in the general field at $z\sim6$. In this way we will be able to address the relative rarity of the J0836 structure in yet another way.  

\subsection{Comparison with other $z\sim6$ structures}
\label{sec:comparison}

In Sect. \ref{sec:3} above we presented statistical and theoretical evidence that the J0836 field appears to be much richer than the average cosmic region at $z\sim6$ based on (1) the photometric overdensity of Z06, (2) the spectroscopic overdensity of B20, and (3) the presence of a very massive companion galaxy. However, we cannot escape the fact that these calculations are a simplification as they do not take into account, for example, the complicated selection functions of candidates and confirmed objects and the small number statistics. One way of addressing these issues has been to use numerical simulations to evaluate the impact of such effects, as selection effects are easily incorporated and simulations offer large volumes from which statistics can be derived \citep[e.g.,][]{overzier09_mr}. On the other hand, however, we should be careful with this approach as well, as many aspects of these simulations remain to be tested by the very observations we try to interpret, and one cannot escape some degree of circular reasoning when comparing the two. Therefore, in this section we will try yet another approach of quantifying the J0836 structure, by comparing with a variety of structures that have been found near other quasars and in the field at $z\sim6-7$. 

\subsubsection{SDSS J1030+0524}
\label{sec:j1030}

Among the luminous quasars at $z\sim6$, there is currently only one other object for which the observations show a certain resemblance to J0836. Quasar SDSS J1030+0524 (J1030) at $z=6.308$ has been among a few, rare quasars with strong evidence for companion structure. The evidence consists of significant overdensities of (photometric) dropout galaxies on various scales \citep{stiavelli05,morselli14,balmaverde17}, and more recently spectroscopic evidence in the form of 4 LBGs and 2 LAEs within a $\Delta z=0.2$ of the quasar redshift \citep{decarli19,mignoli20}. The confirmed LBGs lie at $\sim$8--9\arcmin\ away from the quasar, and at present we have no way of comparing this to the environment of J0836 on these scales. However, \citet{mignoli20} also found two LAEs much closer to J1030, LAE1 at (-4.98,0.14) pMpc and LAE2 at (2.6,0.13) pMpc using our cosmology. Although the proper distances along the l.o.s. are comparable, the projected separations are smaller by a factor of $\sim$2 in the case of the J1030 LAEs. However, it is important to note that they were selected in a very different way using a blind spectroscopic survey over a 1\arcmin$\times$1\arcmin\ field centered on the quasar with VLT/MUSE. The LAEs in the J0836 field were found by targeting photometric dropout objects with slit spectroscopy. The physical properties of the two populations of LAEs are quite different: the J1030 LAEs are  fainter in absolute magnitude ($M_{UV}\gtrsim-20.5$ mag) compared to those in J0836 ($\lesssim$--20.8 mag). The EWs of the two LAEs near J1030 ($EW_0\simeq$11--27 \AA) also appear somewhat lower than the three near J0836 ($EW_0\simeq$10--76 \AA). The relatively high EWs and bright UV magnitudes of the J0836 LAEs are curious because typical LAEs show the opposite trend \citep[e.g.,][]{debarros17}, and could be related to the large quasar proximity zone in which they are situated \citep{bosman20}. We have checked that the LAEs in neither field are affected by \lya\ fluorescence due to the quasar's strong ionizing radiation to a few \% at most \citep[see also][]{bosman20}. Again, we note that the selection techniques used in the two studies were very different. In the case of J0836, current data does not allow us to search for objects as faint as the two LAEs found near J1030. On the other hand, the fact that the J1030 integral field observations did not find any dropout objects with \lya\ as bright as the three objects found near J0836 means that they do not exist near J1030. Summarizing, the comparison shows that both J0836 and J1030 have a large spectroscopic excess of objects relatively close to the quasar (in terms of the projected separation, objects near J1030 are about twice closer than those near J0836), but the physical properties of these objects appear somewhat different (objects near J0836 are brighter in the UV and \lya\ than those near J1030).  

\subsubsection{VIKING J030516.92--315056.0 at $z=6.61$}

\citet{ota18} have detected a large-scale overdensity of star-forming galaxies around the quasar VIKING J030516.92--315056.0 at $z=6.61$. Although several overdense peaks of (candidate) LBGs are found at $\sim$20--40 cMpc away from the quasar, just a single LBG candidate lies within the central 12 arcmin$^2$ equivalent to one HST/ACS pointing. There are also no candidate LAEs in this region with a limiting rest-frame \lya\ $EW_0$ of 15 \AA\ and $L_{Ly\alpha}\gtrsim10^{42.4}$ erg s$^{-1}$, and the surface density of LAEs on larger scales appears to be underdense compared to a control field. There is thus no observational evidence that this particular quasar is surrounded by a structure of galaxies that is similar to that found near J0836, at least not on the relatively small scales probed by our study.   

\subsubsection{Other quasars at $z\sim6$}

A very useful literature overview of searches for structures associated with quasars at $z\simeq5-7$ is given in Table 2 from \citet{mazzucchelli17}. If we limit this list only to quasars for which there exists spectroscopic evidence for associated galaxies on scales similar to that probed by our J0836 study, only 4 out of the original 14 quasars stand out with reported overdensities. Among the $z\sim6$ quasars, this includes only J0836 itself, and J1030 at $z=6.3$ described in detail in Section \ref{sec:j1030} above. The two remaining objects are quasars at $z\approx5$ studied by \citet{husband13}. In one of these, there are 3 objects within $\Delta z=0.11$ in an area similar in scale as the J0836 structure. In the other field, there are 6 confirmed LBGs within $\Delta z=0.06$ (plus an additional quasar), but except for one object these all lie on scales beyond that probed for J0836. Thus, it appears that J0836 and J1030 both represent quite remarkable environments, among all the $z\sim6$ quasars studied to date. 

In addition, \citet{decarli17} derived unique information on the environments of quasars at $z\gtrsim6$ based on the occurrence of companion objects identified based on the \ciialma\ emission line \citep[see also][for similar findings at $z\sim5$]{trakhtenbrot17}. In a survey of 25 quasars, four quasars had close companions (within a projected 600 kpc and 600 km s$^{-1}$). The companion galaxies are not detected in the rest UV, and represent a population of dusty star-forming objects with dynamical masses similar to those of their companion quasars. These massive, close companion objects are consistent with the expected cross-clustering between quasars and star-forming galaxies measured at lower redshifts \citep{garcia-vergara17}. However, the fact that only 4/25 quasars showed \ciialma\ companions, indicates that this type of environment is far from typical. It is important to note that these Atacama Large Millimeter Array (ALMA) observations probe a much smaller field of view than in the other studies described in this paper (survey volume of 400 cMpc$^3$), making any quantitative comparison difficult. Despite it being much further away from the quasar, the stellar mass of the massive companion of J0836 is similar to that of the companion objects identified by \citet{decarli17}. However, no \ciialma\ observations exist for the former.  

\citet{vito21} also found evidence for a quasar at $z=6.5$ involved in a close pair, and 
\citet{yue21} found the first example of a pair of quasars at $z\approx5.7$ (projected separation of $<$10 pkpc). The separations are much smaller than the virial radius of the likely dark matter halos, suggesting that merger-induced triggering could play a role in at least a small fraction of the quasars. 

\subsubsection{COSMOS AzTEC-3 structure at $z=5.3$}

\citet{capak11} found strong clustering near the source COSMOS AzTEC-3 at $z=5.299$ consisting of a spectroscopically confirmed dropout object at $z=5.295$ and several photometric dropouts within a 2 cMpc (0.3 pMpc) radius \citep[see also][]{riechers14}. A quasar at $z=5.30$ lies further away (at 13 cMpc). Although a detailed quantitative comparison is difficult because of the different selections and spectroscopic completeness involved, there appear to be strong similarities between J0836 and this structure in terms of the photometric and spectroscopic overdensities. 

\subsubsection{SXDF protocluster at $z=5.7$}
\label{sec:h19}

\begin{figure}[t]
\centering
\includegraphics[width=\columnwidth]{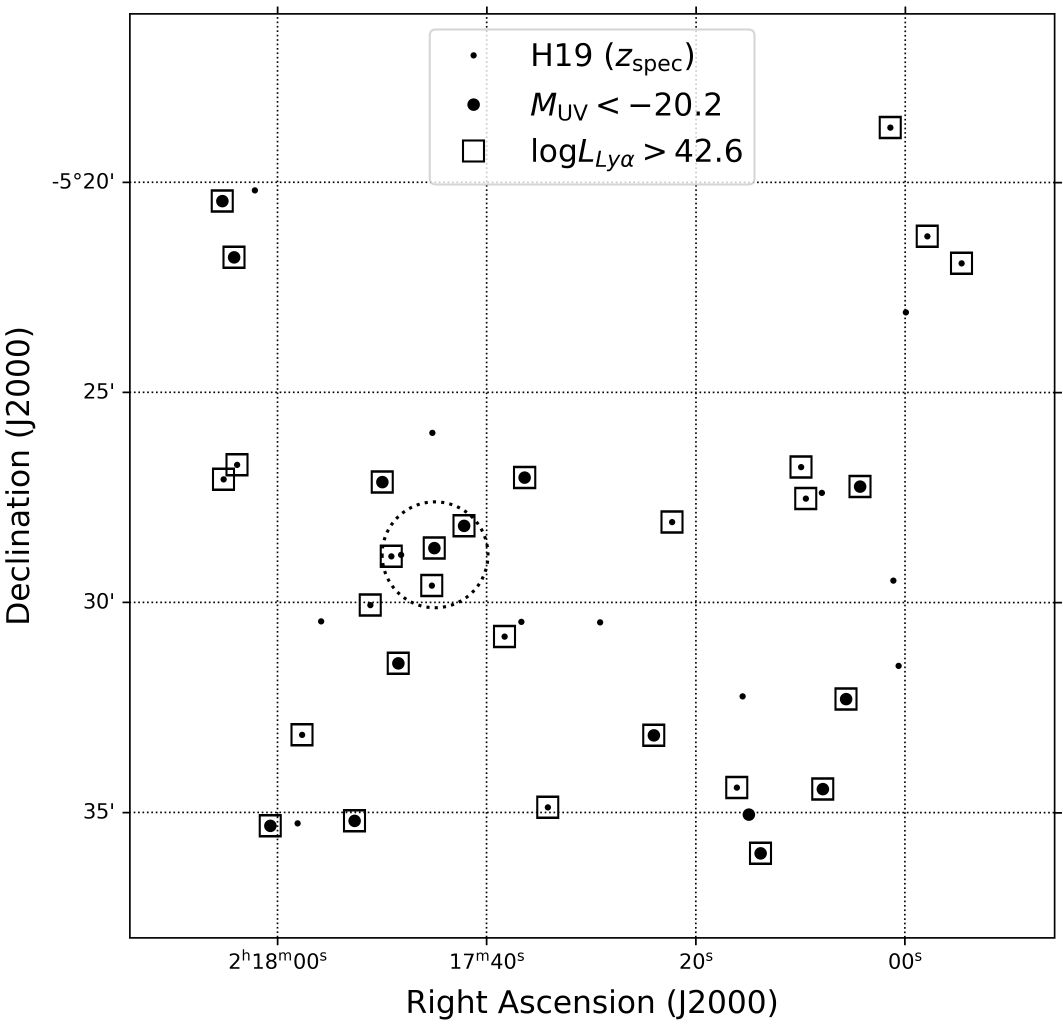}
\caption{Map of the large-scale structure z57OD at $z\approx5.7$ in the SXDF from \citet{harikane19}. The circle with a radius of 1\farcm26 (radius of 0.45 pMpc) that maximizes the number of z57OD galaxies similar to those in the J0836 field has been indicated. There are two objects that would plausibly pass the same selection criteria as those in the J0836 field (large black circles). See Sect. \ref{sec:h19} for details.\label{fig:h19}}
\end{figure}

The Subaru XMM-Newton Deep Field (SXDF) hosts one of the densest structures known at $z\sim5.7$ \citep{ouchi05,jiang18,shibuya18,higuchi19,harikane19}. The structure labeled z57OD corresponds to an overdensity of LAEs with overdensity $\delta=11.5$ and significance of 7.2$\sigma$ when defined using a cylindrical region of 10 cMpc radius and 40 cMpc depth. The core of the structure is characterized by a narrow range in redshift ($\Delta z \approx0.02$), but can be seen in maps with $\Delta z \approx0.12$. In Fig. \ref{fig:h19} we show the z57OD structure using the spectroscopic data from \citet{harikane19}. Objects having absolute rest UV magnitudes brighter than -20.2 mag are indicated with large circles, and objects having \lya\ luminosities larger than 10$^{42.6}$ erg s$^{-1}$ are marked by red squares. The large dotted circle marks a circular 5 arcmin$^2$ region that maximizes the number of known structure members, containing 4 members. Because the SXDF and J0836 structures are, respectively, narrow-band and dropout selected, it is difficult to make a direct quantitative comparison. However, the 4 objects encircled in Fig. \ref{fig:h19} all have \lya\ luminosities at least as high as the 3 confirmed objects near J0836, and 2 of these have UV luminosities similar to those near J0836. Thus, at least on these scales, the J0836 and z57OD structures appear quite similar.   

\subsubsection{HUDF structure at $z\approx5.9$}
\label{sec:m05}

\begin{figure}[t]
\centering
\includegraphics[width=\columnwidth]{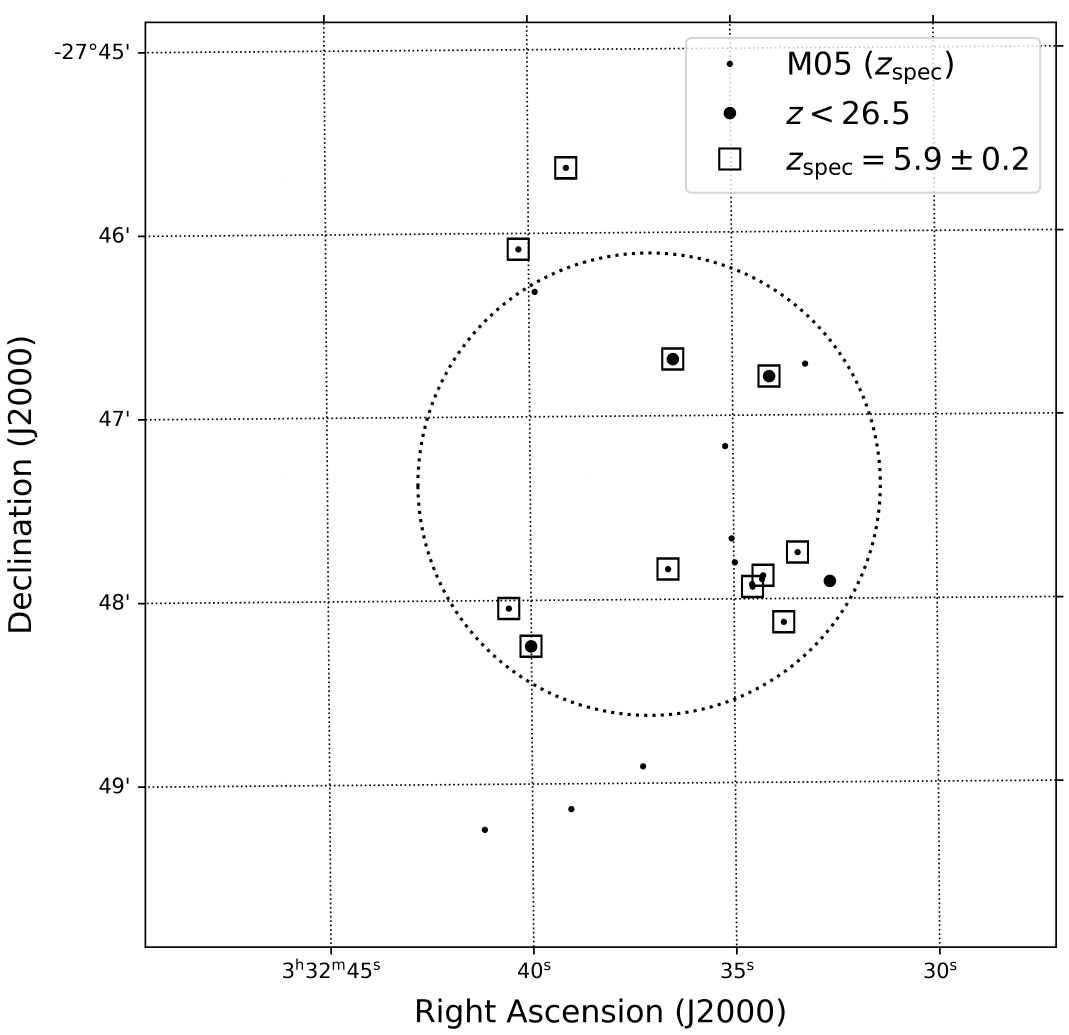}
\caption{Map of the large-scale structure at $z\sim5.9$ in the HUDF from \citet{malhotra05}. The circle with a radius of 1\farcm26 (radius of 0.45 pMpc) that maximizes the number of galaxies has been indicated. There are four objects that would plausibly pass the same selection criteria as those in the J0836 field (large circles), but they have a wider redshift distribution. See Sect. \ref{sec:m05} for details. \label{fig:m05}}
\end{figure}

\citet{malhotra05} identified a structure of \ip-dropout galaxies at $z=5.9\pm0.2$ in the HUDF. They followed up 29 (\ip--\zp$)\ge$0.9 objects with the ACS grism, finding 23 $z\sim6$ objects of which 15 are part of an overdensity of at least a factor of 2 along the line of sight. In Fig. \ref{fig:m05} we show the HUDF structure, indicating objects with (\ip--\zp$)\ge$1.3 (small circles) and those brighter than $z=26.5$ mag (large circles). Objects within the $z=5.9\pm0.2$ redshift spike are marked red. The large dotted circle marks a circular 5 arcmin$^2$ region that maximizes the number of structure members. In this region, there are 7 galaxies, with 3 being brighter than $z=26.5$ mag. The grism redshifts have an accuracy of $\Delta z\approx0.15$, so we will only be able to compare the two structures within $\Delta z=0.2$. The 3 objects thus appear to lie in a redshift interval that is at least as wide as that found for the confirmed objects around J0836. More importantly, the spectroscopic completeness of the \citet{malhotra05} sample is much higher than for the J0836 sample (about 62\% versus 29\%). The J0836 structure thus appears significantly more overdense compared to the HUDF structure, at least at the magnitude limits considered. 

\subsubsection{SDF protocluster at $z=6.01$}
\label{sec:sdf}

\begin{figure}[t]
\centering
\includegraphics[width=\columnwidth]{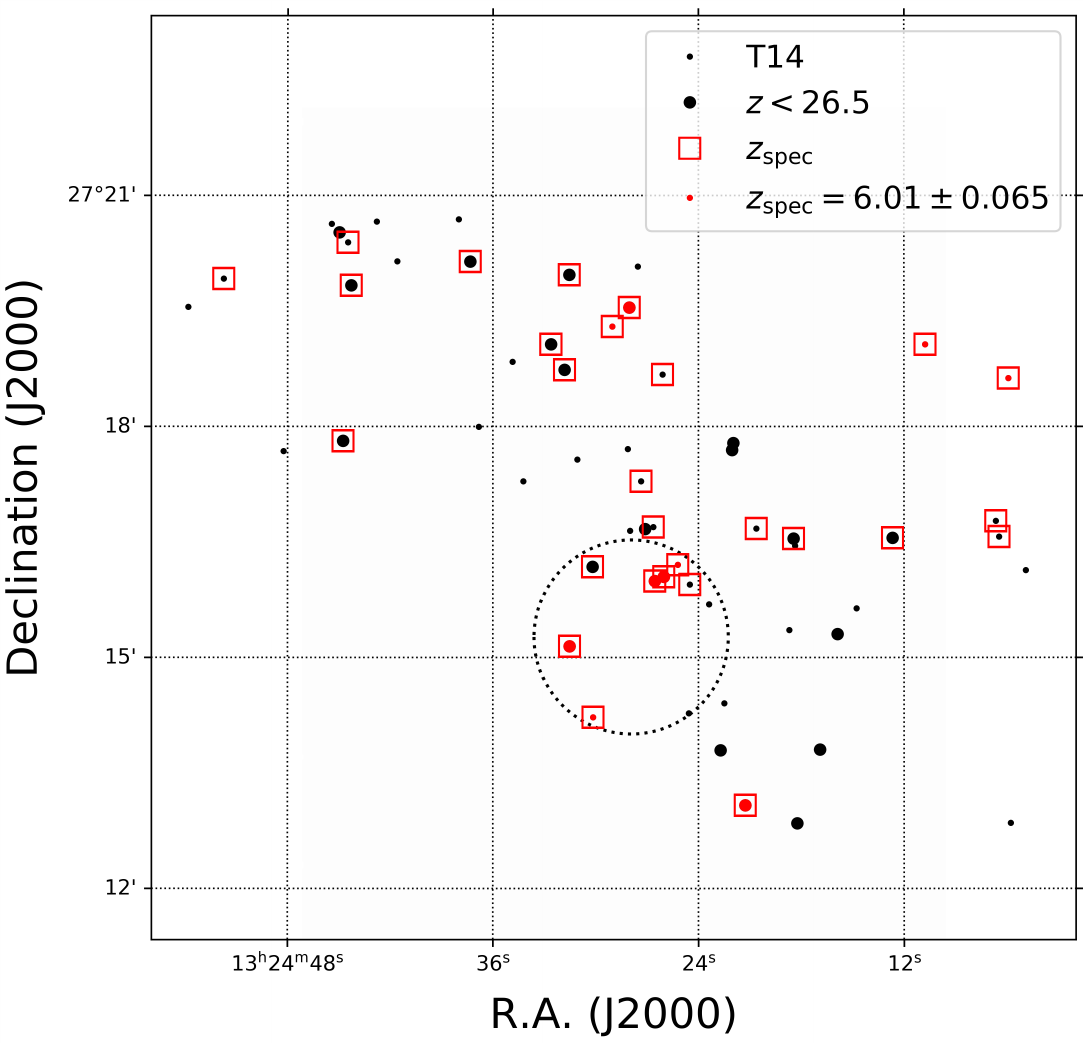}
\caption{Map of the large-scale structure at $z\approx6.01$ in the SDF from \citet{toshikawa12,toshikawa14}. The circle with a radius of 1\farcm26 (radius of 0.45 pMpc) that maximizes the number of galaxies in the SDF structure similar to those in the J0836 field has been indicated. There are three objects that would plausibly pass the same selection criteria as those in the J0836 field (large red circles). See Sect. \ref{sec:sdf} for details.\label{fig:sdf}}
\end{figure}

Among the most impressive large-scale structures known at $z\sim6$ is the protocluster discovered by \citet{toshikawa12,toshikawa14} in the Subaru Deep Field (SDF). Among 28 spectroscopically confirmed $i$-dropout objects at $z\sim6$ lies a structure of 10 objects at $z=5.984-6.047$ ($\Delta z=0.06$) and measuring $20\times20$ cMpc on the sky. The selection of these objects is similar to the dropout selections of Z06 and B20, and the spectroscopic followup similar to that performed by B20 in the J0836 field. In Fig. \ref{fig:sdf} we show the SDF structure, indicating which objects have spectroscopic redshifts (red squares), which objects are brighter than $z=26.5$ mag (large circles), and which objects lie within $\Delta z=0.13$ of the $z=6.01$ redshift spike. The large dotted circle marks a circular 5 arcmin$^2$ region that maximizes the number of structure members. In this region, there are 5 member galaxies, with 3 being brighter than $z=26.5$ mag. This indicates that the J0836 field resembles the densest part of the \citet{toshikawa14} structure.  

\subsubsection{CFHTLS structures at $z\sim6$}

\citet{toshikawa16} performed a color selection of $i$-dropouts in the CFHTLS Deep Fields to a depth of $z\sim26.3-26.5$ mag (about $M_{UV}^{\star,z=6}$). They searched for $>4\sigma$ surface overdensities in regions with a radius of 1 pMpc (2.9\arcmin), suitable for the identification of massive cluster progenitors according to numerical simulations and previous studies \citep{toshikawa12,toshikawa14,toshikawa16}. Five regions were found, and the two most significant overdensities (D1ID01 of 6.1$\sigma$ and D3ID01 of 7.6$\sigma$) were targeted by spectroscopy to confirm the dropout candidates. Three redshifts were obtained in D1ID01, with two objects separated by $\Delta z=0.08$ and 53\arcsec\ ($\sim$0.32 pMpc) and a third object not physically associated at much higher redshift. Two redshifts were obtained in D3ID01, with the two objects separated by $\Delta z=0.007$ and 83\arcsec\ ($\sim$0.49 pMpc). There are no quasars known to be associated with either structure. The three confirmed objects in the J0836 field lie within a projected 0.4 pMpc from the quasar (and about 0.1 pMpc from each other) and have $\Delta z=0.13$. Based on the available photometric and spectroscopic evidence, the J0836 structure thus appears at least as densely populated as the two CFHTLS regions, which themselves represent significant overdensities compared to the general field at $z\sim6$. 

\subsubsection{LAGER-z7OD1 protocluster at $z\approx7$}
\label{sec:hu21}

\begin{figure}[t]
\centering
\includegraphics[width=\columnwidth]{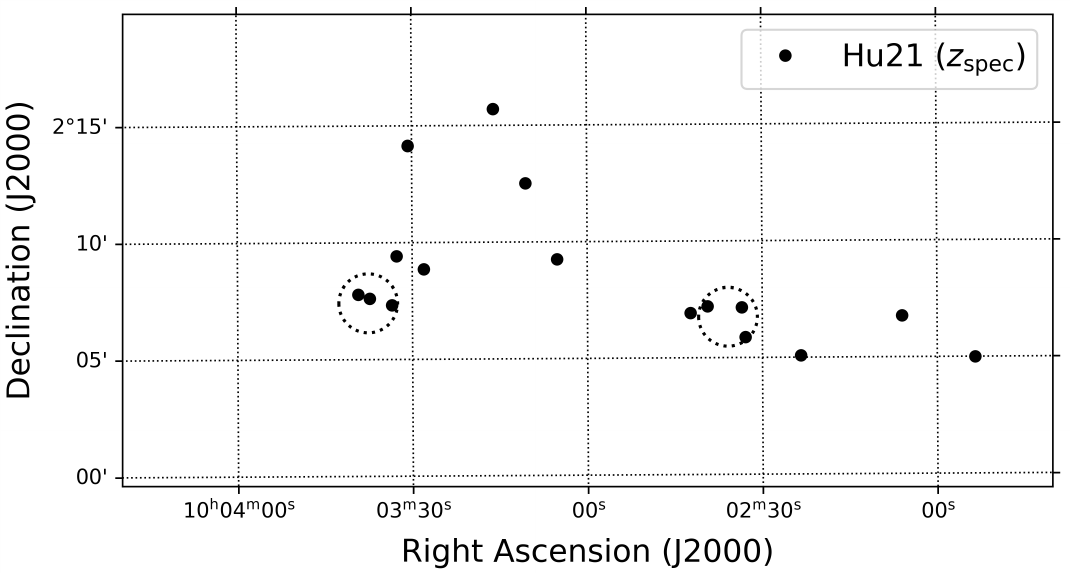}
\caption{Map of the large-scale structure at $z\approx7$ associated with the LAGER-z7OD1 protocluster from \citet{hu21}. Two circles with a radius of 1\farcm26 that maximize the number of galaxies similar to those in the J0836 field have been indicated. There are two locations, each with three objects, that would plausibly pass the same selection criteria as those in the J0836 field. See Sect. \ref{sec:hu21} for details.\label{fig:hu21}}
\end{figure}

\citet{hu21} presented a large overdensity of LAEs at $z\approx7.0$, LAGER-z7OD1. The LAEs were narrow-band selected and have minimum \lya\ luminosities comparable to the LAEs near J0836 (
$\mathrm{log_{10}}L_{Ly\alpha}/(\mathrm{erg~s^{-1}})\gtrsim42.6$). They find 16 spectroscopic LAEs clustered in a region $66\times30\times26$ cMpc$^3$, where the last dimension corresponds to the redshift depth of $\Delta z = 0.072$. Drawing again random circular regions with a radius of 1\farcm26, the maximum number of LAEs encountered is 3, centered around two peaks in the sky distribution of LAGER-z7OD1 (see Fig. \ref{fig:hu21}). This shows that on these scales the structure found around J0836 is not too different from the densest peaks within this large-scale overdense region at $z\approx7$.   

\subsubsection{SC4K survey of LAEs at $z\approx5.7$}
\label{sec:sc4k}

\begin{figure}[t]
\centering
\includegraphics[width=\columnwidth]{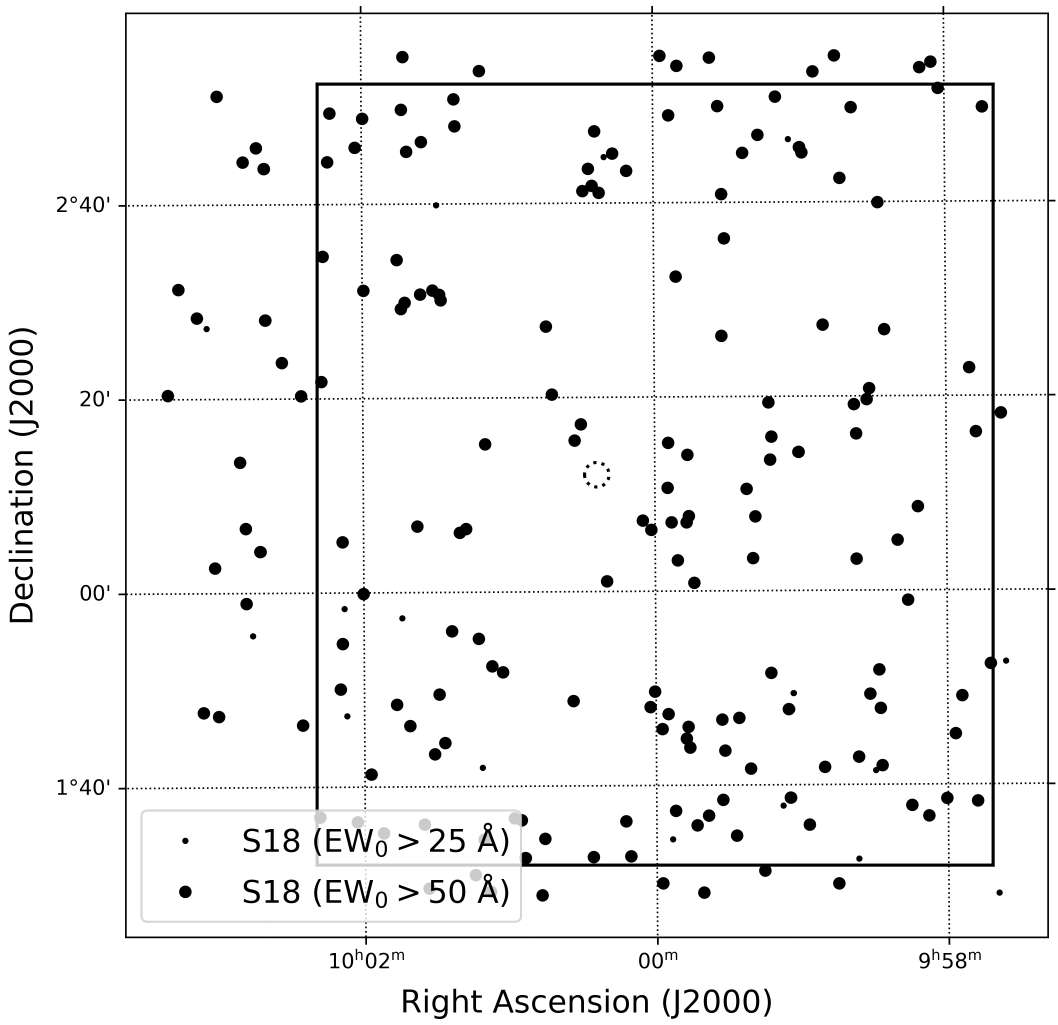}
\caption{Map of the large-scale structure of LAEs at $z\approx5.7$ in the SC4K survey from \citet{sobral18}. LAEs with rest-frame EW $>$25 \AA\ ($>$50 \AA) are indicated by the small (large) circles. A circular region with a radius of 1\farcm26 (0.45 pMpc) similar to the area studied in the J0836 field has been indicated for reference (dashed circle). The area within the large box was used for a counts in cells analysis. The chance of encountering two (three) LAEs that could plausibly pass the same selection criteria as those in the J0836 field amounts to 1\% (0.1\%).  
See Sect. \ref{sec:sc4k} for details. \label{fig:sc4k}}
\end{figure}

To assess the random chance of finding a given number of LAEs in the field, we used data from the $\sim$2 deg$^2$ SC4K survey \citep{sobral18}. We select all narrow-band excess objects detected in the NB816 filter, which is sensitive to \lya\ at $z=5.7\pm0.05$ (FWHM), and thus comparable to the redshift width of the objects in the J0836 field ($\Delta z=0.13$). The contamination by foreground objects in this sample is estimated at about 15\%. It is important to note that the J0836 sample was not narrow-band selected, and the comparison is thus somewhat skewed. 
However, the LAEs in the SC4K survey have redshifts, \lya\ luminosities ($\gtrsim10^{42.6}$ erg s$^{-1}$) and rest-frame EWs ($\gtrsim$50 \AA) comparable to the dropout objects confirmed in the J0836 field by B20. The SC4K survey can thus be used as a conservative reference field for estimating the clustering statistics of these objects (in other words, if we were to perform a survey like SC4K on the J0836 field, we are likely to find even more objects than currently selected, and thus the SC4K reference gives the maximum expectation). In Fig. \ref{fig:sc4k} we show the sky distribution of LAEs from the SC4K survey. We performed a counts-in-cells analysis using a circular region with radius of 1\farcm26 similar to that used in the J0836 field. The chance of randomly finding 2 LAEs in such a small area is at most 1\%, and the chance of finding 3 LAEs is 0.1\% (note that these numbers are likely skewed high due to the contamination of about 15\% and the fact that the J0836 objects were not narrow-band selected). We can conclude from this analysis that the J0836 field contains a number of LAEs that is at least as large as only the densest location in the whole $\sim$2 deg$^2$ SC4K survey.

\subsubsection{Summary of the comparisons}

Based on the various comparisons with known $z\sim6-7$ structures given above, it appears justified to conclude that the J0836 field is comparably rich as the peaks in other rare large-scale structures found at these redshifts. J0836 shows some characteristics of other quasar fields, such as the overdensity of star-forming galaxies found near SDSS J1030+0524 at $z=6.3$ \citep{mignoli20} and the evidence of massive companion galaxies found near some quasars \citep{decarli17,yue21,vito21}, although the scales are very different for the latter. Compared to the non-quasar fields, the J0836 field appears as rich as some of the densest regions known in the field to date. However, the latter include many spectroscopically confirmed surface overdensities extending over angular scales that are much larger than we can probe with the J0836 data, and it is not known whether the similarities between J0836 and these fields would hold on those scales as well. 

\section{Summary and discussion}

\subsection{Summary}

We have shown that the luminous radio-loud quasar J0836 at $z=5.8$ is likely part of a relatively rich structure of galaxies in the early universe. The evidence consists mainly of a photometric overdensity previously found by Z06, a spectroscopic overdensity identified by B20, and the presence of at least one very massive companion galaxy detected at 3.6 $\mu$m with Spitzer/IRAC by \citet{overzier09_irac} allowing a stellar mass estimate. Based on a cross-correlation analysis of galaxies and quasars, we showed that the presence of these companion objects can plausibly be explained by an overdense environment associated with the quasar. We compared the global properties of the J0836 structure with those found near other quasars and in the field at similar redshifts, concluding that the structure resembles the densest peaks in the cosmic density field currently known at $z\sim6-7$, at least at similar depths and scales as probed by our observations. 

This study is significant for a number of reasons. First, it provides new evidence that the first luminous quasars are associated with relatively dense peaks in the cosmic density field. Second, the J0836 structure appears relatively unique among the population of $z\sim6$ quasars studied to date, with one system showing a similar propensity of clustered companion galaxies \citep{mignoli20}, and several other quasars showing evidence for direct interactions on much smaller scales \citep[e.g.,][]{decarli17,vito21,yue21}. Third, although the J0836 structure appears, in some aspects, similar to some of the most clustered regions of star-forming galaxies known at $z\sim6-7$, these regions are clearly not unique to the quasars given that overdensities of galaxies have been discovered in the field as well, sometimes larger and more significant than what has been found near any quasar to date \citep[e.g.,][]{toshikawa14,harikane19,hu21}. 

By combining these different clues, we will discuss a number of important open questions related to the formation of SMBHs, quasar environments and radio jets. 

\subsection{Insights into seed formation}

One of the most important questions in the study of galaxies relates to the origin of the SMBHs. This question is particularly pertinent to the population of luminous quasars at $z\gtrsim6$, given that the masses of their SMBHs managed to rival that of M87 \citep[$\mathcal{M_\mathrm{BH}}=6.5\pm0.2\pm0.7\times10^9$ $M_\odot$;][]{eht19} within 1 Gyr of the Big Bang. This notion has led to a substantial effort in theoretical modeling of possible seed black hole populations and their accretion histories with stellar mass and intermediate mass BHs emerging as the main candidates for the seeds \citep[for reviews see, e.g.,][]{bromm11,inayoshi20,volonteri21}. 

The slightly longer cosmic times available to quasars at $z\lesssim6$ make it significantly ``easier'' to go from a 100 $M_\odot$ seed to a SMBH of a few times $10^9$ $M_\odot$ compared to the quasars at $z\gtrsim7$ \citep[e.g.,][]{banados18,marinello20,pacucci21}. For example, assuming constant accretion with Eddington ratios of 0.8--1.3 onto a seed formed at $z_f=20-10$ is able to produce a SMBH like that in J0836 by $z=5.82$, while the quasar ULAS J1342+0928 at $z=7.54$ requires the existence of a $>$1000 $M_\odot$ seed as early as $z=45$ assuming standard Eddington rate accretion \citep{banados18}. There are at least two problems with the latter scenario: (1) it is not clear if such massive seeds could exist that early, and (2) it is not clear if such an efficient accretion rate could be sustained over such a long cosmic time. 

In scenarios where the growth of SMBHs is allowed to start with an intermediate mass BH (IMBH) seed, the constraints on formation redshift, accretion rates and growth times are significantly relaxed. The biggest uncertainty with these scenarios, however, is whether there exist a viable channel for IMBH formation at early times. Models and simulations have identified a number of ways in which IMBH seeds can form, for example through the merging of numerous stellar mass seeds or through the collapse of massive gas clouds into a DCBH seed (see Sect. \ref{sec:intro}). 

The results presented in this paper cannot constrain any of these scenarios, but the association of J0836 with a relatively dense cosmic structure is very interesting in light of a number of recent model predictions for DCBH formation. The various DCBH scenarios proposed all have in common that they require the quasar to form in an overdense region. The overdense region ensures that the primordial gas cloud is heated through an enhanced local Lyman-Werner photon flux and a high rate of merging (mini)halos \citep[e.g.,][]{wise19,lupi21}. After the collapse, the overdense environment can further ensure a steady accretion rate to form a SMBH and power a quasar.   

The simulations of \citet{wise19} specifically point to a scenario in which the presence of an overdense region of star-forming galaxies several tens of kpc away stimulates the formation of a DCBH seed. When combined with sufficient dynamical heating of atomic cooling halos (ACHs) that are rapidly growing, the nearby star-forming complex only needs to  provide a fraction of the LW flux assumed by earlier models. They find a density of DCBHs in overdense regions at $z=15$ of $10^{-4}-10^{-3}$ cMpc$^{-3}$ with a global number density of $10^{-7}-10^{-6}$ cMpc$^{-3}$ given the rarity of the overdense regions simulated by \citet{wise19}. For normal and void regions of the universe, the number density of DCBH candidates is predicted to be lower by 3--6 orders of magnitude compared to the number density in overdense regions. It is important to point out that the overdense region chosen by \citet{wise19} corresponds to a dark matter halo of a few times $10^{10}$ $M_\odot$ at $z=6$, which is still two orders of magnitude below the typical halo mass of the luminous quasars studied here. If overdense environments indeed enhance the number density of DCBHs by orders of magnitude, perhaps the proto-J0836 environment would make an ideal site. 

The galaxies detected as part of the J0836 structure are unlikely the same as the ones that provided the required LW background at early times. For instance, if we take our model fit results for object A-Z06/A-B20 from Sect. \ref{sec:obja} ($\mathrm{log}_{10}(\mathcal{M}_\star/M_\odot)=10.34_{-0.16}^{+0.30}$ with a mass-weighted age of $0.27_{-0.14}^{+0.28}$ Gyr), we can see that even this massive galaxy likely was not around at $z_f\gtrsim10.5$. Its distance from J0836 is also much too great to have provided a significant Lyman-Werner intensity near the inferred site of seed formation (several Mpc instead of tens of kpc). However, the J0836 structure suggests that an overdense region that predates the formation of the quasar was probably present from early times. It is difficult to be more quantitative at this point. Deep observations with ALMA and the James Webb Space Telescope (JWST) could be used to quantify quasar environments on smaller scales and down to fainter galaxies of much lower bias, and these could then be compared to tailored simulations of DCBH collapse as performed by \citet{wise19}.  

\citet{lupi21} revisit the idea of synchronized pairs (SPs) of ACHs \citep{dijkstra08,visbal14,regan17}, where two halos separated by $<$1 kpc and star formation commencing in one halo $<$5 Myr prior to that in the other provides a high LW flux to drive the formation of a DCBH. They compare this scenario to the dynamical heating (DH) driven collapse similar to that discussed by \citet{wise19}, but focusing specifically on the progenitor of a $3\times10^{12}$ $M_\odot$ at $z=6$. Although the dense environment stimulates the formation of SPs, the pristine halos of the pairs are also easily polluted by metals due to the enhanced supernova activity in the region. These polluted halos are assumed to cool and fragment, and no longer considered candidates for DCBH formation. Comparing this with the DH scenario in the same overdense region shows that, because of the clustering of LW sources in the overdense region, several DH seeds may be formed for each SP seed, although the initial mass of the DH seeds may be lower than the SP seeds. The studies by \citet{wise19} and \citet{lupi21} thus point in the same direction that overdense regions in the early universe may be a necessary ingredient for the formation of DCBHs. 

One could argue that in principle each quasar at $z\gtrsim6$ must have formed in such an overdense environment given their high halo masses \citep{chen21}. However, the finding of relatively rich environments around quasars such as J0836 and J1030 compared to other quasars must then imply that the conditions required for DCBH formation as suggested by theory may have been particularly well met for the progenitor regions of these sources.    

\subsection{Relevance of the radio jets} 

The fast growth rates of the SMBHs powering the $z\sim6$ quasars implies that we are seeing them at a time of potentially significant impact on the growth of the host galaxies through Active Galactic Nucleus (AGN) feedback. The (small) subset of quasars that are radio-loud are the first sources in which we could study, in principle, the interaction between powerful radio jets and the forming interstellar medium. Even though the fraction of quasars that is radio-loud is relatively small ($\sim$10\%, independent of redshift, see \citet{banados15}), new radio surveys at low frequencies detect a significant fraction of the population of radio-quiet quasars at $z>5$ \citep{gloudemans21}. There is growing evidence that at least at low redshift faint radio structures present also in radio-quiet quasars may be disproportionally responsible for the feedback observed \citep[e.g.,][]{jarvis21}.

At these redshifts, there are still relatively few radio sources known. J0836 appears as the 8th most distant radio-loud AGN in the overview of \citet[][their Table 6]{banados21}. While optically the most luminous, it has the lowest radio-loudness parameter ($R_{2500}=16\pm1$). We determine a bolometric luminosity of $L_\mathrm{bol}=5.3\pm0.02 \times10^{47}$ erg s$^{-1}$ using \citet{richards06}, and with a black hole mass $\mathrm{log}_{10}(\mathcal{M}_\mathrm{BH}/M_\odot)=9.48\pm0.55$, we find an Eddington ratio of $L_\mathrm{bol}/L_\mathrm{Edd}=1.4_{-1.0}^{+3.5}$. This range indicates that J0836 lies along the upper envelope of Eddington ratios found for high redshift quasars compiled by \citet{banados21}.    

The compact radio size \citep{frey05}, steep spectral index and evidence for a peaked radio spectrum \citep{wolf21} suggest that J0836 is not beamed and that it is a young radio source in which the jets are still confined to the inner kpc scales. With a (projected) size of 40 pc and assuming a typical hot spot advancement speed of ($0.2\pm0.1$) $h^{-1}$ $c$ \citep{giroletti09}, the kinematic age in the young source scenario is in the range $\approx300-1000$ yr. If the jets have an inclination angle as small as 1\degr\ along the line of sight (still large enough such that it does not become a blazar), the radio age could be as high as $\approx3\times10^4$ yr. This is much shorter than the most recent quasar phase of $\sim2\times10^{5-7}$ yr estimated by B20. Given these relatively short jet life-times compared to the black hole and quasar accretion time scales inferred for J0836, it appears unlikely that the (current) radio jets have had any meaningful impact on the growth of the SMBH. Alternatively, the radio jets could be recurrent, or the radio source is much older than inferred from its maximum linear size, as expected in the case of jets that are confined by a dense ISM \citep[e.g.,][]{odea21}. 

What does seem relevant, however, is the fact that J0836 represents yet another case of a high redshift radio-loud AGN associated with a relatively dense environment \citep[e.g.,][]{overzier06a,venemans07,hatch14,overzier16}. The ability of SMBHs to form strong radio jets is generally assumed to scale as a function of black hole mass, accretion rate and spin \citep{blandford77,fanidakis11}, and all three parameters may be expected to be enhanced in overdense regions due to the enhanced gas accretion rates and merger activity. On the other hand, radio jets may be the consequence of energy extracted from the accretion disk, in which case the SMBH grows much faster due to higher accretion rates \citep[][]{blandford82}. A recent discussion of this relevant to the jets in radio-loud quasars at $z\sim6$ is given by \citet{connor21}.

Quasar J0836 is different from the radio-loud quasar PSO J352.4034-15.3373 at a very similar redshift \citep{banados18_rlqso,rojas21,connor21}, mainly in the sense of its very compact linear size \citep[$\sim$40 pc versus $\sim$1.6 kpc;][]{frey05,momjian18}, radio power \citep[rest-frame $L_{\nu,1.4}\sim7\times10^{26}$ W Hz$^{-1}$ versus $\sim5\times10^{27}$ W Hz$^{-1}$;][]{wolf21,banados18_rlqso}, and radio loudness parameter \citep[$16\pm1$ versus $1470_{-100}^{+110}$;][]{banados21,rojas21}. It also appears to be quite different from the distant radio galaxy TGSS J1530+1049 at $z=5.72$ discovered by \citet{saxena18} with  rest-frame $L_{\nu,1.4}=1.6\times10^{28}$ W Hz$^{-1}$ and a radio morphology consisting of two radio components with a linear extent of $\sim$2.5 kpc \citet{gabanyi18}. However, all these sources are consistent with being relatively young radio sources in which the radio structures are comparable or within the scale of the host galaxies. These sources thus offer an excellent opportunity for investigating the interaction between radio jets, galaxy formation and SMBH growth.   

\subsection{Concluding remarks}

The answer to the question of what were the seeds of today’s SMBHs is coming within our reach. Likely it will require the synthesis of data and clues from a wide variety of independent upcoming experiments. Gravitational wave astronomy has already begun to map the black hole mass gap \citep{ligo20}, and the proposed Laser Interferometer Space Antenna (LISA) mission \citep{lisa17} and Einstein Telescope \citep{et10,et20} will significantly widen the mass and distance range of detectable black hole mergers, thereby addressing directly the mass distribution of the population as a function of redshift. JWST could in principle detect massive DCBH seeds -- if they exist -- through their unique spectral signatures in the infrared, provided they are not too rare \citep{pacucci16,natarajan17,woods19,woods21a}. The Nancy Grace Roman Space Telescope, due to its much wider field of view, could perform a search employing gravitational lensing \citep{whalen20}. An upper limit on the number density of DCBH candidates could already be used to constrain the DCBH formation scenario, and perhaps shift the focus of modeling efforts to seeds originating from Population III stars instead. In any case, JWST will also map for the first time in detail the amount of black hole activity in typical galaxies during the first billion years, and the masses of SMBHs in (faint) quasars, thereby constraining the total black hole demographics and indirectly the seed population. Evidence for Population III stars either at high redshift or in the Local Group as well as advanced models could constrain the range of the masses of this first generation of stars and thus the mass range that their end-products could realistically achieve \citep[e.g.][]{hirano14,placco21,woods21a,woods21b}. Searches for IMBHs in globular clusters and dwarf galaxies will provide yet another unique constraint on the seed population \citep{greene12,mezcua17,latimer21}. The search for distant obscured black holes will also greatly benefit from the new capacity in the X-rays provided by the Athena or Lynx missions \citep{pacucci15,amarantidis19}. Extremely Large Telescopes will then be needed to confirm and characterize the sources found. Yet another window on the first stages of black hole formation will be provided at radio wavelengths. Deep radio surveys with the Square Kilometer Array may detect DCBHs through their core radio emission \citep{whalen21}, and also find the most distant objects in the universe capable of hosting radio jets, the smoking gun of massive, spinning black holes.

In the more immediate future, it will be extremely interesting to see (1) to what extend the compact radio structures seen in the most distant radio galaxies and quasars interact with the ISM of their young, forming host galaxies, (2) if there is any evidence for environmental triggers of this radio activity, and (3) to establish definitively what are the typical environments of the radio-loud and radio-quiet AGN at this important epoch. Several programs have been scheduled on the HST and the JWST that aim to do just that within the next two years (e.g., HST programs 16258/16693 and JWST programs 1205/1554/1764/2028/2078). 

\begin{acknowledgments}
The author is grateful to Catarina Aydar, Sarah Bosman, Roberto Decarli, Benedikt Diemer, Pavel Kroupa, Marco Mignoli, Sof\'ia Rojas, Aayush Saxena, Kazuhiro Shimasaku and Mauro Stefanon for comments or answering questions during the preparation of this manuscript. The author thanks the anonymous referee for insightful comments and recommendations. The author was supported in this work by a productivity grant (302981/2019-5) from the National Council for Scientific and Technological Development (CNPq).
\end{acknowledgments}


\vspace{5mm}
\facilities{HST(ACS), Spitzer(IRAC)}


\software{astropy \citep{astropy13,astropy18}, 
            CosmoCalc \citep{wright06},
            BagPipes \citep{carnall18},
          %Cloudy \citep{2013RMxAA..49..137F}, 
          Source Extractor \citep{bertin96}, 
          Colossus \citep{diemer18}
          }




\bibliography{ms}{}
\bibliographystyle{aasjournal}



\end{document}